\documentclass[reprint, amsmath, amssymb, aps, pre, floatfix, showkeys, superscriptaddress]{revtex4-1}

\usepackage{graphicx}
\usepackage{dcolumn}
\usepackage{bm}
\usepackage{color}
\usepackage{tikz}
\usepackage{pgfplots}
\usepackage{mathtools}
\usepackage{array}
\usepackage{appendix}

\pgfplotsset{compat=1.18}

\DeclareMathOperator*{\argmin}{arg\,min}
\newcommand*\mean[1]{\bar{#1}}

\begin{document}

\title{Data-driven guessing and gluing of unstable periodic orbits}

\author{Pierre Beck}
\email{pierre.beck@epfl.ch}
\address{
Emergent Complexity in Physical Systems Laboratory (ECPS), \'Ecole Polytechnique F\'ed\'erale de Lausanne, 1015 Lausanne, Switzerland
}

\author{Jeremy P. Parker}
\email{jparker002@dundee.ac.uk}
\address{
Division of Mathematics, University of Dundee, Dundee DD1 4HN, United Kingdom
}

\author{Tobias M. Schneider}
\email{tobias.schneider@epfl.ch}
\address{
Emergent Complexity in Physical Systems Laboratory (ECPS), \'Ecole Polytechnique F\'ed\'erale de Lausanne, 1015 Lausanne, Switzerland
}

\date{\today}

\begin{abstract}
Unstable periodic orbits (UPOs) are believed to be the underlying dynamical structures of spatio-temporal chaos and turbulence. Finding these UPOs is however notoriously difficult. Matrix-free loop convergence algorithms deform entire space-time fields (loops) until they satisfy the evolution equations. Initial guesses for these robust variational convergence algorithms are thus periodic space-time fields in a high-dimensional state space, rendering their generation highly challenging. Usually guesses are generated with recurrency methods, which are most suited to shorter and more stable periodic orbits. Here we propose an alternative, data-driven method for generating initial guesses, enabled by the periodic nature of the guesses for loop convergence algorithms: while the dimension of the space used to discretize fluid flows is prohibitively large to construct suitable initial guesses, the dissipative dynamics will collapse onto a chaotic attractor of far lower dimension. We use an autoencoder to obtain a low-dimensional representation of the discretized physical space for the one-dimensional Kuramoto-Sivashinksy equation, in chaotic and hyperchaotic regimes. In this low-dimensional latent space, we construct loops based on the latent POD modes with random periodic coefficients, which are then decoded to physical space and used as initial guesses. These loops are found to be realistic initial guesses and, together with variational convergence algorithms, these guesses help us to quickly converge to UPOs. We further attempt to `glue' known UPOs in the latent space to create guesses for longer ones. This gluing procedure is successful and points towards a hierarchy of UPOs where longer UPOs shadow sequences of shorter ones.
\end{abstract}

\maketitle

\section{Introduction}

It is widely accepted that unstable periodic orbits (UPOs) play an important role in supporting chaotic dynamics in many driven dissipative nonlinear systems \cite{devaney1986, chaosbook}. 
Periodic orbits are believed to be dense in the chaotic attractor of such systems \cite{devaney1986, Viswanath2003} and are organized in a hierarchical fashion \cite{chaosbook, Christiansen1997, BudanurShort2017}, where longer orbits shadow a sequence of shorter ones. 
A well-known example displaying this hierarchical organization structure is the chaotic Lorenz ODE (ordinary differential equation) system with its famous chaotic attractor, which resembles the shape of a butterfly. Periodic orbits and chaotic trajectories shadowing them can be encoded by their sequential passage from one wing to the other. This leads to a description in terms of symbolic dynamics and reveals that periodic orbits are related to each other, with longer ones shadowing shorter ones \cite{Viswanath2003}. The symbolic encoding, moreover, allows one to enumerate all UPOs. 

For chaotic PDEs, including the Navier-Stokes equations in the turbulent regime, we expect the same hierarchical organization of UPOs as in low-dimensional ODEs; and formal periodic orbit theory aimed at describing ergodic averages in terms of expansions over UPOs or, alternatively, over prime cycle sequences characterizing those UPOs, assumes it \cite{chaosbook, Cvitanovic1995}. However, at least in the context of fluid flows, we are not aware of any direct demonstration of the hierarchical UPO organization. This fact highlights the algorithmic and numerical challenges inherent in identifying UPOs in high-dimensional chaotic systems. These computational challenges are especially severe for long periodic orbits that may be shadowing several shorter ones.

Physically relevant nonlinear dissipative chaotic PDEs, including model equations and the full Navier-Stokes equations, are usually formulated in 1-3 spatial dimensions, but their solution space is a formally infinite-dimensional function space. The spatial and temporal dimensions are typically discretized with many points, yielding a high-dimensional set of coupled ODEs. The identification of UPOs of the Navier-Stokes equations \cite{Kawahara2001} suggests that a similar dynamical systems approach as in ODEs can be applied to PDEs, with spatiotemporal chaos being viewed as a chaotic walk between invariant solutions, such as equilibria, UPOs and invariant tori \cite{Parker2022a, Parker2023a, Lan2006NewtonsTori}. 
However, for 3D fluid flows, only very few UPOs have been identified due to the difficulties in computing them efficiently. Specifically, an envisioned hierarchical organization of UPOs has not been directly and conclusively demonstrated thus far.  

Invariant solutions have contributed to the understanding of flow physics \cite{KawaharaAnnualRev}, such as self-sustaining near-wall coherent structures \cite{Kawahara2001, Waleffe2001, Waleffe2003}, the self-organizing oblique stripe patterns in plane Couette flow \cite{Reetz2019}, or recovering turbulence statistics \cite{Kawahara2001,VANVEEN2006}. Moreover, invariant solutions have been observed in experimental settings \cite{Hof2004, GrigorievPRL2017}, and as a result, much research has been conducted on the identification of such invariant solutions. In particular, for UPOs, this is usually done in two steps: first by defining an adequate guess for a UPO and secondly by converging this guess to a solution of the system. The obvious and most common methods for finding UPOs are Newton shooting methods \cite{Viswanath2007, Viswanath2009, Cvitanovic2010, Chandler2013}. Here, the initial guess for the UPO is represented by a point in state space, namely the initial condition, and a period $T$. These two are then varied until the time-integrated trajectory closes in on itself. Typically, the optimization step is solved with a Newton algorithm. However, the exponential error amplification encountered when time-integrating a chaotic dynamical system leads to convergence issues, particularly when searching for long UPOs. Multi-shooting methods \cite{sanchez2009, BudanurShort2017, sajjadthesis} attempt to control the exponential error amplification by dividing the integration interval into smaller sub-intervals, however they still rely on time-integrating a chaotic system and as a consequence do not have very robust convergence properties.

More recently, loop convergence algorithms \cite{Lan2004} and their matrix-free variations \cite{Azimi2022, Parker2022} have shown to be effective in finding UPOs \cite{Cvitanovic2010a, Lasagna2017}. The guess is now a space-time field that is already periodic (a loop) with time $T$ but it does not satisfy the flow equations. The matrix-free variational methods from \cite{Azimi2022, Parker2022} deform the loop until its tangent vectors align everywhere with the flow vectors prescribed by the equations. This removes the time-integration aspect and consequently the challenges associated with an exponential error amplification characteristic of chaotic systems.  

In order to converge to UPOs (and in particular many distinct and long ones) we require a method to construct good guesses. Conventionally, guesses are extracted from recurrency methods \cite{Kawahara2001, Viswanath2007, Cvitanovic2010, Chandler2013, Morita2010, Lathrop1989}, where one looks for sub-trajectories in a long DNS of the system that almost close in on themselves. The downside of this method is that a trajectory is required to follow a UPO for an entire period, which is unlikely due to their unstable nature. As a result, this method is biased towards the same few frequently visited UPOs (usually short and less unstable ones). Although short periodic orbits are expected to have larger contributions in periodic orbit theory \cite{chaosbook}, longer and more unstable periodic orbits are still necessary to obtain more accurate statistics \cite{Chandler2013, Page2022}. They are also interesting from a control-theoretic point of view as they capture the dynamics and can be tracked for varying control parameter values \cite{Lasagna2017}. Moreover, some dynamics appear to only be captured by long periodic orbits. \citet{Lan2008} study the Kuramoto-Sivashinsky PDE, and for their control parameter of choice, the shortest UPO they find has period 12.08, while orbits that connect dynamically different parts of the chaotic attractor have periods of around 355.34 or more. Identifying guesses for such long UPOs that converge in the context of shooting methods is extremely difficult and requires unrealistically precise and extremely rarely observed recurrences. However, loop-based convergence algorithms that formulate a guess as a loop representing an entire space-time field have much more robust convergence properties than shooting approaches \cite{Lan2004, Lan2008}, in the sense that they can also converge from inaccurate guesses \cite{Azimi2022}. This robustness together with the periodic nature of guesses in loop convergence algorithms may allow to define guesses within the algorithm's convergence radius using alternative methods instead of a recurrency analysis. 

While the solution space is formally infinite-dimensional, trajectories in nonlinear, chaotic, driven, dissipative systems (such as the Navier-Stokes equations) collapse on a chaotic attractor once transients have died down \cite{Hopf1948}. This attractor can be embedded in a curved manifold of far lower dimension - often termed the inertial manifold \cite{Temam1990}. Consequently, the high-dimensionality of the system's state space that renders the identification of UPOs so challenging, may be interpreted as an artifact of not knowing the most appropriate coordinates for describing the lower-dimensional intrinsic dynamics within the inertial manifold. If one had access to coordinates that approximately parametrized the lower dimensional (as compared to the discretization dimension) manifold that the attractor is embedded in, one could use these reduced coordinates to construct initial guesses for UPOs. In combination with the robust loop-convergence algorithms, even randomly drawn closed curves that lie in the inertial manifold and match the statistical properties of the attractor may be sufficient to define realistic guesses and identify UPOs. In analogy to the analysis of low-dimensional ODEs such as the Lorenz system, concatenating sequences of short UPOs to formulate guesses for long UPOs within such reduced coordinates may further allow to construct the hierachical sets of UPOs for PDEs that are expected to exist in theory, but so far have not been demonstrated directly. 

Various methods exist for obtaining approximations of manifold coordinates. Linear model order reduction methods such as Dynamic Mode Decomposition (DMD) \cite{Schmid2010} and Principal Component Analysis (PCA) \cite{Pearson1901LIII.Space/i,Hotelling1936} (more commonly known as Proper Orthogonal Decomposition (POD) in the fluid dynamics community) are very popular for dimensionality reduction. While they capture a great deal of information, they are known to generalise less well to highly nonlinear systems and are outperformed by nonlinear methods, such as spectral submanifolds \cite{Haller2016, Cenedese2022} or autoencoders from deep learning \cite{Goodfellow2016, Linot2023a}. The fact that autoencoders in particular seem efficient in giving a low-dimensional representation of spatiotemporal chaos is demonstrated by \citet{PageBrenner2020}. They train an autoencoder to identify low-dimensional embeddings of monochromatically forced Kolmogorov flow. They find that even for very low latent dimensions, such as $N_h = 3$, they obtain small losses, and come to the conclusion that much of the dynamics, such as low-dissipation events, live in a low-dimensional space. Even high-dissipation events are captured by only slightly larger latent dimensions, such as $N_h = 32$. Within the low-dimensional latent space, \citet{PageHoley2023} re-define a recurrence function to obtain guesses for periodic orbits. \citet{Linot2022Data-drivenEquations, Linot2019, Zeng2024} also explore autoencoders and combine them with a neural network in the latent space for time-series prediction. In particular, they study how the quality of the autoencoder improves as the latent dimension approaches the manifold dimension of the chaotic attractor.  

We propose a new method for guessing periodic orbits in the context of variational loop convergence algorithms by randomly drawing loops that are in statistical agreement with the attractor. To this end, we define the loops as linear combinations of the POD modes with random periodic coefficients chosen to match the moments of the flow. We can arbitrarily adjust the complexity and length of these loop guesses and target longer UPOs by increasing the number of `twists' or `turns' in the loop. As a pre-processing step, we use an autoencoder to obtain an approximation of the low-dimensional manifold coordinates and generate our guesses inside the autoencoder's latent space. We will discuss how including the autoencoder impacts the performance of the guesses. Moreover, inside the low-dimensional latent space, we attempt to concatenate or `glue' orbits together. By gluing orbits, we define longer and more accurate guesses in a hierarchical fashion as has been observed for ODEs but not for PDEs, to the best of our knowledge.

The structure of this paper is as follows: in section \ref{sec:background} we give a brief reminder of loop-based convergence methods and introduce our setup for illustrating the methods in the Kuramoto-Sivashinsky PDE in chaotic and hyperchaotic regimes. In section \ref{sec:methods} we describe our methods used for defining initial guesses and describe the complete convergence setup for finding UPOs. We also define the notion of latent gluing and explain how we generate new, longer guesses by concatenating shorter UPOs. In section \ref{sec:application} we first apply these methods to Kuramoto-Sivashinsky for parameter value $L = 39$ (low-dimensional chaos), and then in the hyperchaotic case at $L = 100$. We discuss and conclude in section \ref{sec:conclusion}.

\section{Background}
\label{sec:background}

\subsection{Loop convergence methods}
\label{sec:looping}
\citet{Azimi2022} treat the general PDE ${\partial_t\bold{u} = \boldsymbol{F}(\bold{u})}$ for a real field $\bold{u}(\bold{x},t)$ on a $n$-dimensional spatial domain $\mathcal{X} \subset \mathbb{R}^n$ with initial condition $\bold{u}_0$. The flow function $\boldsymbol{f}^t$ advances the dynamical system in time ${\bold{u}(t) = \boldsymbol{f}^t(\bold{u}_0) = \bold{u}_0 + \int_0^t\boldsymbol{F}dt'}$ where $t$ is the time. A fixed point $\bold{u}^*$ is a solution that satisfies ${\boldsymbol{F}(\bold{u}^*) = 0}$. A periodic orbit is characterised by an instantaneous field $\bold{u}$ (the initial condition) and a period $T>0$ that satisfy ${\boldsymbol{f}^T(\bold{u}) - \bold{u} = \boldsymbol{0}}$ such that for any $T^* < T$ this equation is not satisfied.

In shooting methods, a guess for a periodic orbit consists of an initial condition $\bold{u}_0$ and a period $T$. The pair $(\bold{u}_0, T)$ are then varied until the time-integrated curve closes in on itself. This is done by solving the equation ${\boldsymbol{f}^T(\bold{u}_0) - \bold{u}_0 = \boldsymbol{0}}$, typically via Krylov subspace methods \cite{Kelley2003, Sanchez2004} including Newton GMRES hook-step methods \cite{Viswanath2007, Viswanath2009, Cvitanovic2010} and variations \cite{Dennis1996, Chandler2013, Duguet2008a}. More recently, \citet{Page2022} define a cost function that is the norm of this equation and use gradient-based optimization to minimize it until a root is found.

In the loop convergence algorithm of \citet{Azimi2022}, the guess consists of an entire space-time field $\bold{u}(\bold{x}, t)$, defined on $\mathcal{X}\times [0,T)_{periodic}$ that is time-periodic with a guess period $T$, but does not necessarily satisfy the evolution equations of the system. A priori, the period $T$ is unknown, and hence the field is re-scaled such that $s = t / T$ and $\bold{\tilde u}(\bold{x}, s) := \bold{u}(\bold{x}, sT)$. Hence $\bold{\tilde u}$ is a function defined on $\mathcal{X}\times [0,1)_{periodic}$ mapping to $\mathbb{R}^n$. A solution $\bold{\tilde v}(\bold{x}, s)$ of the system then satisfies the re-scaled equation

\begin{equation}
    \frac{1}{T} \frac{\partial \bold{\tilde v}}{\partial s} = \boldsymbol{F}(\bold{\tilde v})
\end{equation}

Defining the residual vector $\bold{r}$ for a loop $\bold{\tilde u}(\boldsymbol{x}, s)$ by

\begin{equation}
    \boldsymbol{r} = \boldsymbol{F}(\bold{\tilde u}) - \frac{1}{T} \frac{\partial \bold{\tilde u}}{\partial s}
\end{equation}

we obtain the cost function $J$ of a loop:

\begin{equation}
    J := \int_0^1\int_\mathcal{X}\boldsymbol{r} \cdot \boldsymbol{r}\;d\boldsymbol{r}ds
\end{equation}

A loop with $J = 0$ satisfies the flow equations and is a closed curve, and hence a periodic orbit. Conceptually, the loop is deformed until it satisfies the flow equations. Geometrically, the tangent vectors of the loop are aligned with the velocity vectors of the flow.

\citet{Lan2004} minimize the cost function with a variational approach. While the method is very robust \cite{Lan2008}, it involves a large Jacobian and therefore does not scale to high-dimensional systems, a challenge laid out explicitly by \citet{Fazendeiro2010} and \citet{Boghosian2011}. Inspired by a similar approach on equilibria by \citet{Farazmand2016}, \citet{Azimi2022} formulated a matrix-free adjoint-based method (which we will use in this paper) for minimizing $J$ over the space of loops which scales to high-dimensional systems, like the Navier-Stokes equations. They use the Kuramoto-Sivashinsky equation as a test-bed, while \citet{Parker2022} apply a similar method to 2D Kolmogorov flow and explicitly address incompressibility in different ways. \citet{Ashtari2023b} deal with the challenge of computing pressure in the presence of solid walls in the 3D Navier-Stokes equations and introduce a method to accelerate the adjoint-based variational method with DMD. \citet{Ashtari2023} also show that a similar approach can be used to compute connecting orbits through the construction of an analogous cost-function that undergoes an adjoint-based minimization process.

\subsection{The Kuramoto-Sivashinsky equation}
\label{sec:KS}
The 1D Kuramoto-Sivashinsky equation (KSE) is a nonlinear PDE which arises in the modelling of the evolution of viscous liquid films down a vertical plane \cite{Sivashinsky1980}, reaction-diffusion systems \cite{Kuramoto1976}, and flame-fronts \cite{Sivashinsky1977} and exhibits chaotic behaviour for certain parameter values. In this paper, we use the following non-dimensional formulation of the KSE:

\begin{equation}
    u_t + uu_x + u_{xx}  +u_{xxxx} = 0
\end{equation}
where $u$ is the velocity field and we assume that the spatial domain is $L$-periodic, such that $u(x,t) = u(x+L,t)$. The domain length $L$ is the control parameter. The system is invariant under spatial and temporal translation, as well as under reflection $x\to-x$, $u\to-u$. We will work in the anti-symmetric subspace $u(x) = -u(-x)$ (denoted by $\mathbb{U}^+$ in \cite{Cvitanovic2010a}), as was done for example in \citet{Lan2008} and \citet{Lasagna2017}, and is enforced by keeping the real part of the Fourier coefficients equal to 0. This discretizes the spatial translation invariance, and reduces it to $x \to x + L/2$. If this symmetry was not imposed, the spatial translation invariance could also be filtered out by the method of slices and recovered via its reconstruction equations \cite{Rowley2000, Marensi2023, Engel2023}. The system is also invariant under Galilean transformations, however this is filtered out by the imposed anti-symmetry condition. 

\subsubsection{Low-dimensional chaos : $L = 39$}
\begin{figure}
    \centering
    \includegraphics[width = \columnwidth]{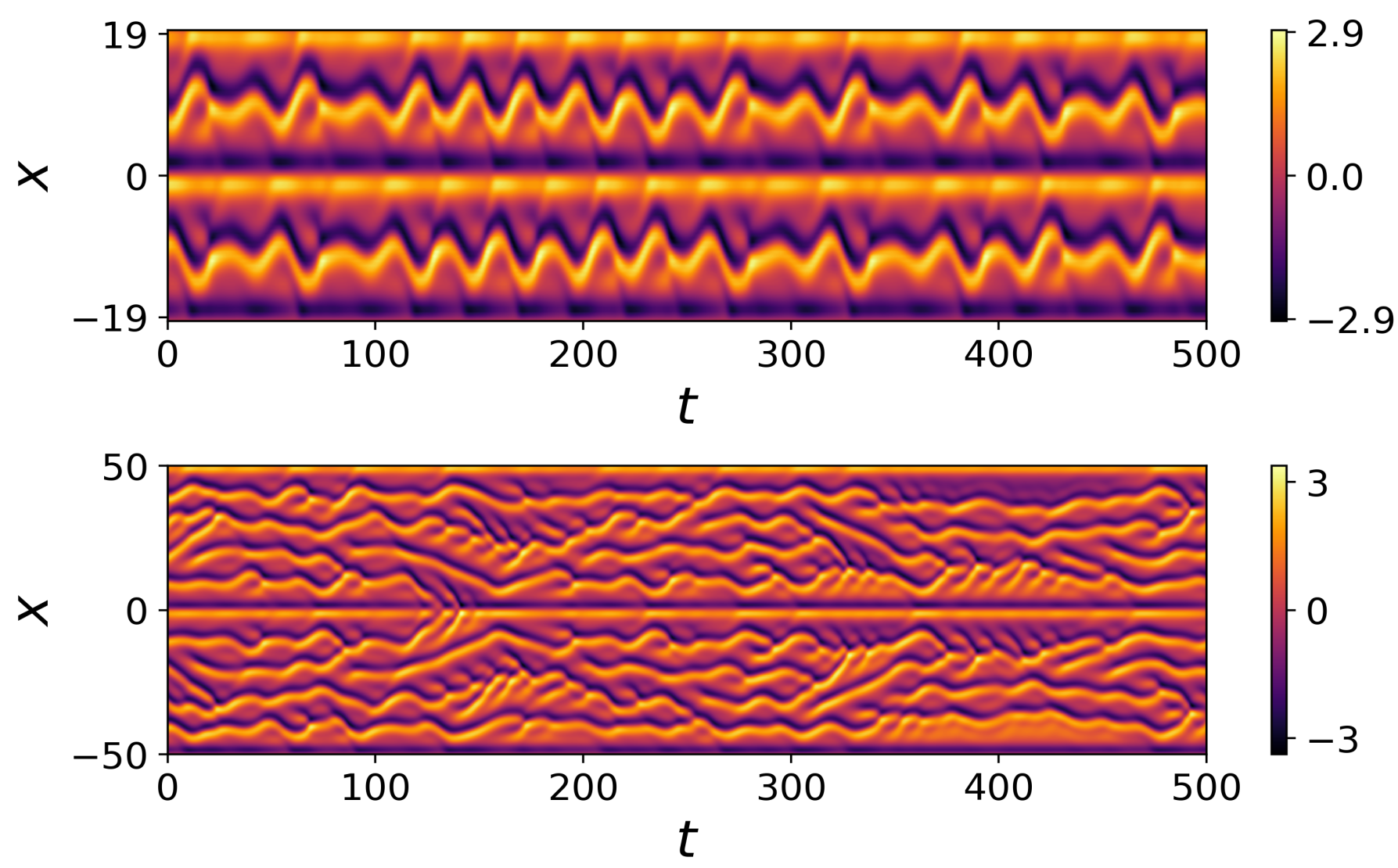}
    \caption{Space-time plot of example trajectories of the Kuramoto-Sivashinsky equation with $L = 39$ (top) and $L = 100$ (bottom) in the anti-symmetric subspace, discretized by $N_x = 64$ and $N_x = 170$ spatial nodes respectively.}
    \label{fig:phys_traj}
\end{figure}

Initially, we set $L = 39$, as in \cite{Lasagna2017}, for which low-dimensional chaos is observed. Although this is not the simplest chaos observed for the KSE \cite{Edson2019, Wilczak2020,Abadie2024}, it is simple enough for us to show-case our methods before we move to a more complicated system. In this case, we discretize the spatial dimension with $N_x = 64$ points, turning the scalar function $u$ into a 64-dimensional state vector $\boldsymbol{u}$ (not to be confused with the $n$-dimensional continuous field $\bold{u}$ in section \ref{sec:looping}). The data of a long trajectory ($dt = 0.1, T_{max} = 155,000$) is generated using the ETDRK4 scheme \cite{Kassam2005}. To ensure that the data is indeed just from the chaotic attractor, we cut off the first 5,000 time units. A space-time plot of an example trajectory generated in this setup is shown at the top of figure \ref{fig:phys_traj}.

\subsubsection{Hyperchaos : $L = 100$}
In the second instance, we set $L = 100$, when the system is hyperchaotic, meaning there are 2 or more positive Lyapunov exponents. According to \citet{Edson2019}, there are 5 positive Lyapunov exponents. At $L = 39$, the dynamics are still relatively simple (albeit chaotic) due to the restricted symmetry and the narrow spatial domain. The more complex, hyperchaotic system resembles true spatiotemporal chaos, which is more akin to turbulence in fluids which exhibits multiple positive Lyapunov exponents  \cite{Hassanaly2019}. In this case, we discretize the spatial dimension with $N_x = 170$ points, for which we observe a sufficient drop between the largest and smallest frequency of the time-averaged energy spectrum, while not over-resolving the system and making computations too slow. A typical trajectory of the system is presented at the bottom of figure \ref{fig:phys_traj}, showing the increased complexity. Again, we generate one long trajectory where we cut off the first 5,000 time units, in order to make sure that all our data is part of one chaotic attractor.

\section{Methods}
\label{sec:methods}
In this section we first introduce the data-driven dimensionality reduction technique we use, namely autoencoders, which will serve as a pre-processing step in our guessing procedure. We then explain how we generate loop guesses as linear combinations of POD modes and describe the algorithmic procedure which we employ to converge to periodic orbits. Finally, we set out the gluing procedure to connect two existing periodic orbits, which serves as a guess for longer UPOs and helps us identify a hierarchy of UPOs.

\subsection{Data-driven dimensionality reduction}
\label{sec:data-driven}
\subsubsection{Architecture}
We apply an autoencoder to reduce the physical discretization dimension $N_x$ to a latent dimension $N_h$. Autoencoders are neural networks that consist of two parts, namely the \textit{encoder} $\mathcal{E}: \mathbb{R}^{N_{in}}\rightarrow\mathbb{R}^{N_h}$ and the \textit{decoder} $\mathcal{D}: \mathbb{R}^{N_h}\rightarrow\mathbb{R}^{N_{in}}$. Given input data $\boldsymbol{y}\in \mathbb{R}^{N_{in}}$, we would like to train the network so that approximately $\mathcal{D}\circ\mathcal{E} (\boldsymbol{y}) \approx \boldsymbol{y}$ with $N_h \ll N_{in}$.

The architecture of the autoencoder we employ in this paper varies slightly with the control parameter $L$, but the general architecture is similar: the encoder $\mathcal{E}$ consists of three dense layers, with the third one having $N_h$ nodes. The decoder has the same setup as the encoder, just in reverse. For the $L = 100$ case, we add four linear layers after the third dense layer of $\mathcal{E}$ in order to minimize the rank of the latent space as described by \citet{Jing2020AE} and also explored by \citet{Zeng2024}. 
Illustrations of the networks are shown in Appendix \ref{sec:app_training}. Note that we chose these architectures because they worked well for our purposes, but we do not claim that they are perfectly optimized. In both variations of the KSE explored later, we use ReLU activation functions on all layers except the linear layers: $\mathrm{ReLU}(x) = \max\{0,x\}$.

\subsubsection{Training}
Due to the imposed anti-symmetry in the KSE, the numerical data obtained from direct numerical simulation (as described in \ref{sec:KS})  is also anti-symmetric, rendering half of the components of $\boldsymbol{u}$ superfluous. Moreover, one of the remaining components is always zero, and so we only use $N_{in} = N_x / 2 - 1$ components of $\boldsymbol{u}$ as inputs $\boldsymbol{y}$. These inputs $\boldsymbol{y}$ are re-scaled before they are used for training. We subtract the mean and then use a min-max re-scaling:
\begin{equation}
    \boldsymbol{y}^* = \frac{(\boldsymbol{y} - \boldsymbol{y}_{mean}) - \boldsymbol{y}_{min}}{\boldsymbol{y}_{max}-\boldsymbol{y}_{min}} \in [0,1)^{N_{in}}
\end{equation}
where division is applied component-wise. The minimum and maximum are vectors and also taken component-wise over $\boldsymbol{y} - \boldsymbol{y}_{mean}$. We drop the $^*$ for what follows for convenience. As loss function $\mathcal{L}$, we use the mean relative difference between the input $\boldsymbol{y}$ and the output $\mathcal{D}\circ\mathcal{E}(\boldsymbol{y})$ rather than the standard mean-squared error, as dividing by the norm of $\boldsymbol{y}$ scales the loss in an interpretable manner. For $N$ data points $\{\boldsymbol{y}_n\}_{n=1}^N$, the loss is:

\begin{equation}
    \mathcal{L} = \frac{1}{N}\sum_{n=1}^N\frac{||\mathcal{D}\circ\mathcal{E}(\boldsymbol{y}_n) - \boldsymbol{y}_n||^2}{||\boldsymbol{y}_n||^2}
\end{equation}

Details on the training hyper-parameters are given in Appendix \ref{sec:app_training}. The best choice for $N_h$ is a priori not known other than that we would like $N_h \ll N_{in}$. We take topological quantities such as the Kaplan-Yorke dimension $D_{KY}$ as guidance \cite{Linot2019, Linot2022Data-drivenEquations}. Note, however, that $D_{KY}$ is a global average dimension of the attractor, and locally the topology might be more complicated. Since we are not looking for perfect guesses, $D_{KY}$ is a good starting point. For each parameter choice of $L$ and a selection of values for $N_h$, we train 20 autoencoders with different initializations for each value of $N_h$. Of the 20 networks, we pick the one with the best test loss. We then decide on the value of $N_h$ by considering $D_{KY}$ and by observing the test loss with respect to $N_h$.

In this section, we introduced the data-driven dimensionality reduction techniques that we will use to obtain an approximate low-dimensional representation of the high-dimensional discretized system. In the next section, we explain how the autoencoder's latent space gives us effective low-dimensional coordinates appropriate for defining loop guesses that are time-periodic space-time fields, lie on the chaotic attractor by matching the statistics of the flow and can be adjusted in length.

\subsection{Loops based on POD modes}
\label{sec:pod}
We construct the guesses for UPOs based on linear combinations of the proper orthogonal decomposition (POD) modes \cite{Pearson1901LIII.Space/i, Hotelling1936} with periodic coefficients. In general, consider a long time-series stacked in a matrix $\boldsymbol{Z}\in\mathbb{R}^{p\times N}$, where the rows are $p$ time-steps $\{\boldsymbol{z}_i\}_{i = 1}^p$, and $\boldsymbol{z}_i\in \mathbb{R}^N$ (in our context, $\boldsymbol{z}$ could be $\boldsymbol{u}$, or latent variables $\mathcal{E}(\boldsymbol{y})$). To calculate the POD modes, we compute the covariance matrix of the zero-mean time-series: let $\boldsymbol{\tilde{z}}_i = \boldsymbol{z}_i - \boldsymbol{\mean{z}}$, where $\boldsymbol{\mean{z}}$ is the mean flow, and let $\boldsymbol{\tilde{Z}}\in\mathbb{R}^{p\times N}$ be the corresponding zero-mean time-series. The unbiased estimator $\boldsymbol{C}$ for the covariance matrix is then given by 
\begin{equation}
    \boldsymbol{C} = \frac{1}{p-1} \boldsymbol{\tilde{Z}}^T\boldsymbol{\tilde{Z}}\in \mathbb{R}^{N\times N}
\end{equation}

The POD modes $\boldsymbol{\phi}_1, ..., \boldsymbol{\phi}_N$ are the eigenvectors of $\boldsymbol{C}$
\begin{equation}
    \boldsymbol{C}\boldsymbol{\phi}_k = \lambda_k\boldsymbol{\phi}_k
\end{equation}
with corresponding eigenvalues $\lambda_1, ...,\lambda_N$. Without loss of generality, the modes $\boldsymbol{\phi}_1, ..., \boldsymbol{\phi}_N$ are ordered such that the eigenvalues are in decreasing order ${\lambda_1 \geq ... \geq \lambda_N \geq 0}$ (note that since the covariance matrix $\boldsymbol{C}$ is symmetric and positive semi-definite, its eigenvalues are real and positive).

The POD modes can be interpreted as fluctuations around the mean flow. Thus, we define a loop $\boldsymbol{L}(\boldsymbol{x}, s)$ via a linear combination of the $\boldsymbol{\phi}_k (\boldsymbol{x})$ with periodic coefficients $a_k(s,\{X_{m,k}\}_{m = 0}^M)$
\begin{equation}
    \boldsymbol{L}(\boldsymbol{x}, s) = \boldsymbol{\mean{z}} + \sum_{k = 1}^N a_k(s, \{X_{m,k}\}) \boldsymbol{\phi}_k(\boldsymbol{x})
\end{equation}
where $s$ is a periodic parameter and the $\{X_{m,k}\}_{m = 0}^M$ are a sequence of independent and identically distributed (iid) random variables with a distribution $X_{m,k} \sim X$ to be determined. 
We want the distribution of $X$ to be so that loops on average match the first and second moments of the flow. Matching the first moment then means that the loop-average over the distribution of loops should agree with the mean flow. We denote this averaging by $\mathbb{E}_{X,s}[...]$, meaning $\mathbb{E}_{X}[\langle ... \rangle_s]$, where $\langle ... \rangle_s = \frac{1}{2\pi}\int_0^{2\pi}... \;ds$. This gives the following two moment matching conditions:
\begin{align}
    \mathbb{E}_{X,s}[\boldsymbol{L}] &= \boldsymbol{\mean{z}} \label{eqn:eq_mean}\\
    \mathrm{cov}_{X,s}(\boldsymbol{L}) \coloneqq \boldsymbol{C}^{(\boldsymbol{L})}  &= \boldsymbol{C} \label{eqn:eq_cov}
\end{align}
For our purposes, we assume that the $X_{m,k}$ do not depend on $s$, and thus the order of integration between $\mathbb{E}_X$ and $\langle ... \rangle_s$ does not matter. 
We match the first (equation \ref{eqn:eq_mean}) and second moment (equation \ref{eqn:eq_cov}) by setting 
\begin{align}
    \mathbb{E}_{X,s}[a_k] & = 0 \label{eqn:eq_match1}\\
     \mathrm{var}_{X,s}(a_k) & = \lambda_k \label{eqn:eq_match2}
\end{align}

for $k = 1, ..., N$. The detailed derivation of equations \ref{eqn:eq_match1} and \ref{eqn:eq_match2} is given in Appendix \ref{sec:moments}.

We now define the coefficients $a_k$. Since we want them to be time-periodic, we write the coefficients as a sum of sines and cosines
\begin{equation}
    \label{eqn:sin_cos_main}
    a_k(s, A_{:,k}, B_{:,k}) = \sum_{m = 0}^M \alpha_{m} [A_{m, k} \cos(ms) - B_{m, k} \sin(ms)]
\end{equation}
where $s\in[0,2\pi)$, $M$ is the number of sine/cosine modes to be included in the sum, and the coefficients $A_{m, k}, B_{m, k} \sim X$ are iid. The $\alpha_{m}$ are constants that give different weights of choice to higher frequency terms, for example $\alpha_{m} = (m + 1) / (M + 1)$. 

By substituting equation \ref{eqn:sin_cos_main} into equations \ref{eqn:eq_match1} and \ref{eqn:eq_match2}, we find that
\begin{align}
   \mathbb{E}_X[A_{m,k}] &= \mathbb{E}_X[B_{m,k}] = 0 \label{eqn:coeff1}\\
       \mathrm{var}_X(A_{:,k}) &= \mathrm{var}_X(B_{:,k}) = \lambda_k \bigg( \sum_{m = 0}^M \alpha_{m}^2\bigg)^{-1}\label{eqn:coeff2}
\end{align}

Thus, setting $ A_{:,k}, B_{:,k} \sim \mathcal{N}\Bigl(0, \lambda_k \Bigl( \sum_{m = 0}^M \alpha_{m}^2\Bigl)^{-1}\Bigr)$ fulfills our requirements. Again, the detailed derivation is given in Appendix \ref{sec:coeffs}. 

The parameter $M$ allows us to define longer guesses. Geometrically, adding higher modes to the sum in equation \ref{eqn:sin_cos_main} introduces extra `twists' or `turns' in our loop. This can be seen intuitively by thinking of Poincaré sections: the shortest UPOs are expected to have only $p=1$ intersections with an adequately chosen Poincaré section. UPOs with $p = 2$ intersections would require an extra twist in their geometric loop representation to intersect twice with the hyperplane. Thus, if we want to target UPOs with $p$ Poincaré intersections, we set $M = p$. We will verify this intuition in section \ref{sec:application}.

Defining loop guesses via equation \ref{eqn:sin_cos_main} allows us to generate random guesses by ad-hoc loops that are time-periodic and statistically lie on the attractor by matching moments up to second order. Moreover, the number of sine/cosine modes $M$ allows us to adjust the length of the guess and therefore to directly target long UPOs. Nevertheless, the guesses themselves do not know anything about the dynamics, making it a crude approach for loop definition. While we will see that we can easily generate promising guesses this way and converge to UPOs, we will also present the method's current limitations. 

Finally, the derivation of loops in this section is valid both in the physical space and in the autoencoder's latent space. Therefore, it would not be unreasonable to do the guessing entirely in the physical space, without the autoencoder. However, we found that such guesses based on the first $N_h$ POD modes are outperformed by guesses based on the latent POD modes of a $N_h$-dimensional latent space of an autoencoder, both in convergence success rates but also the variety of UPOs found (we provide detailed numbers on tests we conducted in Appendix \ref{sec:pod_guesses}). This is likely due to the POD modes only giving a linear subspace, which restricts the guesses for more complicated systems. Also in general, POD modes are outperformed by autoencoders for more complicated systems such as fluid flows \cite{Linot2020, Linot2023a}.

\subsection{Algorithm for guessing and converging loops to UPOs}
\label{sec:algorithm}
In sections \ref{sec:data-driven} and \ref{sec:pod}, we laid out the methods for obtaining a low-dimensional representation of the high-dimensional discretized system and for generating guesses that are time-periodic space-time fields that match the flow statistics up to second order. Together with a loop convergence algorithm, we can now devise a general scheme for guessing loops and converging them to periodic orbits:
\begin{enumerate}
    \item Obtain data $\{\boldsymbol{u}_m\}_{m=1}^M$ of the PDE of interest via direct numerical simulation.
    \item Train an autoencoder for an adequate choice of $N_h$ to get $\mathcal{E}$ and $\mathcal{D}$.
    \item Obtain the latent POD modes $\boldsymbol{\xi}_1, ... ,\boldsymbol{\xi}_{K}$ with eigenvalues $\gamma_1, ..., \gamma_{K}$ based on a timeseries $\{\boldsymbol{h}_i\}_{i = 1}^p$, where $\boldsymbol{h}_i\in \mathbb{R}^{N_h}$, and $K \leq N_h$ is such that $\forall i = 1, ..., K$ we have $\gamma_i > 0$.
    \item Define a loop $\boldsymbol{L}$ in the latent space following the approach from section \ref{sec:pod}:
    \begin{equation}
    \label{eqn:loop}
        \boldsymbol{L}(s) = \boldsymbol{\mean{h}} + \sum_{k = 1}^K a_k(s)\boldsymbol{\xi}_k
    \end{equation}
    where 
    \begin{equation}
        a_k(s) = \sum_{m = 0}^M \alpha_{m} [A_{m, k} \cos(ms) - B_{m, k} \sin(ms)]
    \end{equation}
    
    and $s\in [0,2\pi)$. The coefficients $A_{m,k}, B_{m,k}$ are randomly drawn from a normal distribution $\mathcal{N}\Bigl(0, \lambda_k \Bigl( \sum_{m = 0}^M \alpha_{m}^2\Bigl)^{-1}\Bigr)$.

    \item Decode the loop to physical space $\mathcal{D}(\boldsymbol{L})$.
    \item Use the adjoint solver from Azimi et al. \cite{Azimi2022} to converge the cost function $J$ of the loop to ${J\approx10^{-4}}$. Small weight adjustments are made to the gradient of the period $T$ for stability.
    \item Use a Newton solver on the loop residual $J$ to converge the loop to machine precision.
\end{enumerate}
As a post-processing step, every UPO's time-resolution is increased to 256 time-steps and re-converged.

\subsection{Latent gluing of UPOs}
\label{sec:gluing}
Given two periodic orbits $\mathcal{P}_1$ and $\mathcal{P}_2$ with respective periods $T_1, T_2$, we will explore the possibility of concatenating or `gluing' them together in latent space and using this as an initial guess for a longer periodic orbit. The motivation behind this is symbolic dynamics, where a trajectory is described symbolically by its sequential passage through different parts of state space \cite{Viswanath2003, Lan2008}. This appeals to a hierarchy of periodic orbits, where long UPOs shadow shorter ones \cite{chaosbook}. Since this hierarchy appears to exist in ODE systems, we expect that this also applies in the KSE's physical and latent spaces. 

Applying the encoder $\mathcal{E}$ to each time-step of the discretized orbits ${\boldsymbol{P}_1\in\mathbb{R}^{N_{t_1} \times N_x}}$ and ${\boldsymbol{P}_2\in\mathbb{R}^{N_{t_2} \times N_x}}$ gives their latent representations ${\boldsymbol{L}_i\in\mathbb{R}^{N_{t_i} \times N_h}}$. To define the latent glued orbit ${\boldsymbol{G}\in\mathbb{R}^{ (N_{t_1} + N_{t_2}) \times N_x}}$, we start by finding the indices $i^*, j^*$ that minimize the distance between the two orbits in the latent space

\begin{equation}
    i^*, j^* = \argmin_{i,j} ||\boldsymbol{L}_1^{(i)} - \boldsymbol{L}_2^{(j)}||_2
\end{equation}

where $\boldsymbol{L}_1^{(i)}$ and $\boldsymbol{L}_2^{(j)}$ are the $i$-th and $j$-th rows (or time-steps) of $\boldsymbol{L}_1, \boldsymbol{L}_2$ respectively. We define this minimal distance to be
\begin{equation}
    \ell_2 = ||\boldsymbol{L}_1^{(i^*)} - \boldsymbol{L}_2^{(j^*)}||_2
\end{equation}

The number of time-steps of each discretization are adequately chosen such that ${N_{t_1}/N_{t_2}  \approx T_1/T_2}$. Define the naively glued orbit $\boldsymbol{G}_0$ by vertically stacking $\boldsymbol{L}_1$ and $\boldsymbol{L}_2$ at the points of closest passage

\begin{equation}
    \boldsymbol{G}_0 = 
    \begin{pmatrix}
        & \boldsymbol{L}_1^{(1:i^*)} \\
        & \boldsymbol{L}_2^{((j^*+1):end)} \\
        & \boldsymbol{L}_2^{(1:j^*)} \\
        & \boldsymbol{L}_1^{((i^*+1):end)}
    \end{pmatrix}
\end{equation}

Since this introduces a jump discontinuity, we smooth $\boldsymbol{G}_0$ in the latent space to get the new guess $\boldsymbol{G}$: We set the high-frequency temporal modes in Fourier space to zero and keep only the lowest 1/6 positive (and lowest 1/6 negative) frequencies. The guess is then defined as $\mathcal{D}(\boldsymbol{G})$ (where $\mathcal{D}$ is applied to each time-step) with guess period ${T = T_1 + T_2}$, and then follows steps 6 and 7 of algorithm \ref{sec:algorithm}.

\section{Results}
\label{sec:application}
We apply the methods described in section \ref{sec:methods} to the KSE for two different parameter regimes: first $L=39$, for which low-dimensional chaos is observed, and then for the hyperchaotic case at $L = 100$.
\subsection{Low-dimensional chaos $L = 39$}
\begin{figure}
    \centering
    \includegraphics[width = \columnwidth]{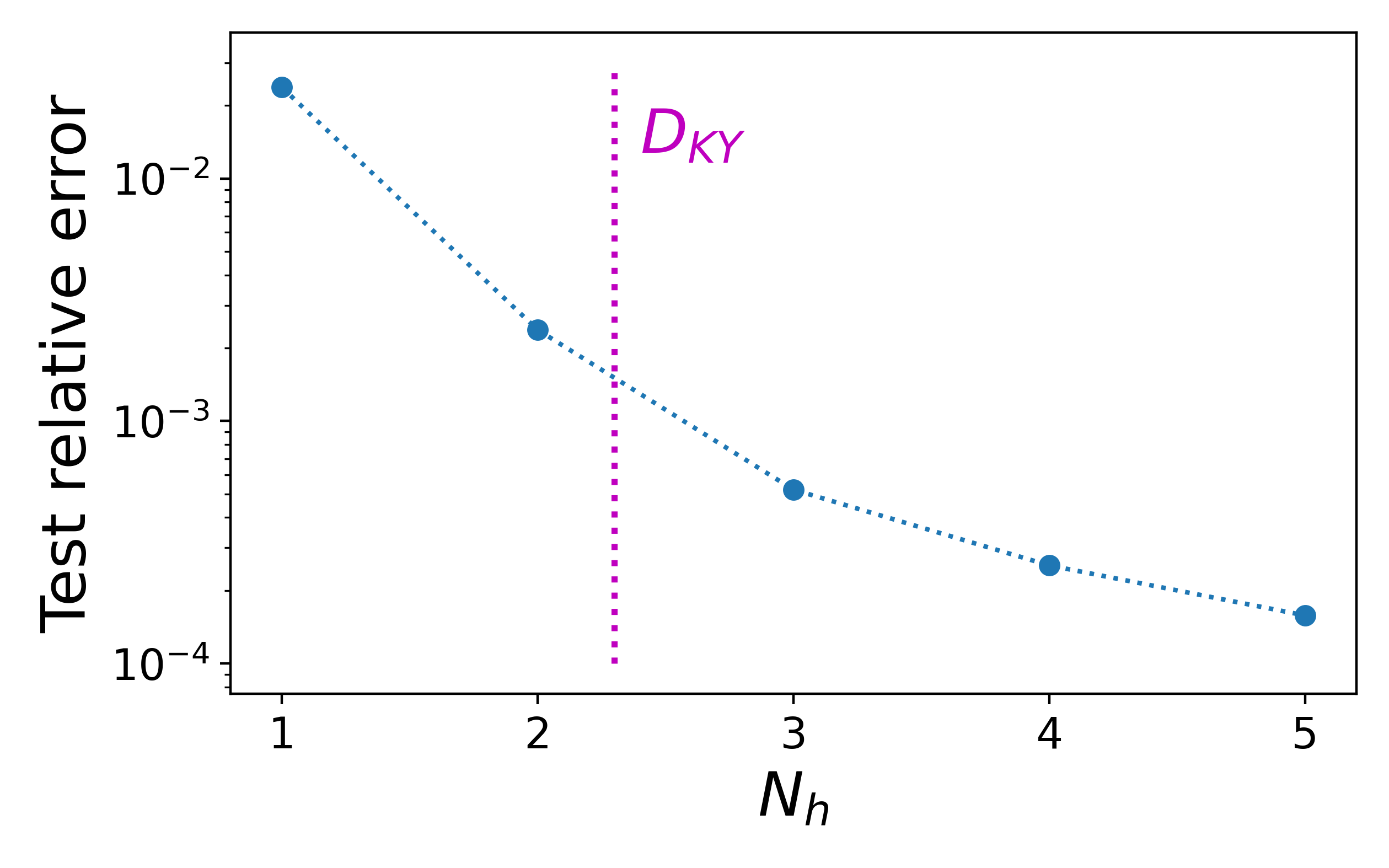}
    \caption{The final best test losses of the 20 neural networks trained for each latent dimension $N_h = 1$ to $N_h = 5$, with $D_{KY} \approx 2.3$ indicated in magenta. For this system, we end up working with $N_h=  3$.}
    \label{fig:L39_network_losses}
\end{figure}

\begin{figure}
    \centering
    \includegraphics[width = \columnwidth]{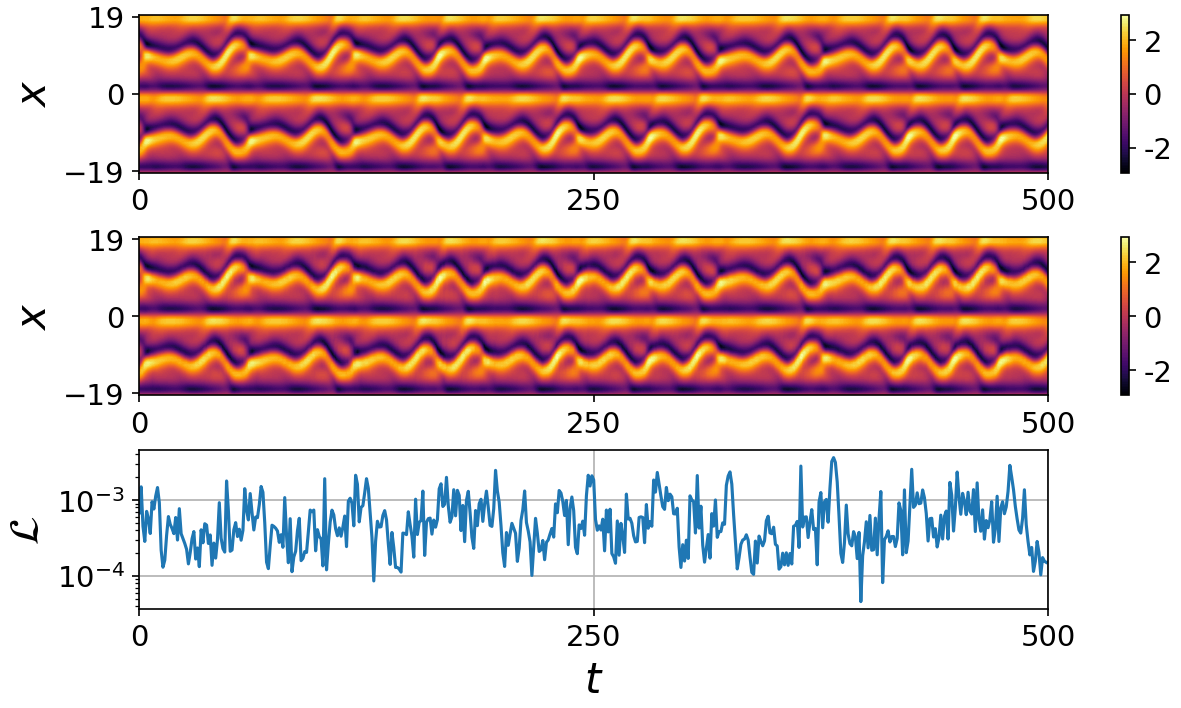}
    \caption{Top: Physical trajectory of the KSE in the test set. This data was not used in training. Middle: Output when applying the autoencoder with latent dimension $N_h = 3$ to the trajectory. Bottom: Loss $\mathcal{L}$ over time.}
    \label{fig:test_set}
\end{figure}

Due to the imposed anti-symmetry in the system, half of the $N_x = 64$ discretization components are redundant. One of the remaining components is always zero. Thus the input vectors for the autoencoder have $N_{in} = 31$ dimensions. We train 20 autoencoders for each of $N_h = 1,...,5$ and find the final test losses shown in figure \ref{fig:L39_network_losses}. As expected, the loss decreases when we increase $N_h$. The Kaplan-Yorke dimension for this system is $D_{KY}\approx 2.3$ \cite{Edson2019}. We decide to continue working with $N_h = 3$, as we observe a reasonable loss value and it is in agreement with $D_{KY}$. This choice has the nice bonus that we can visualize our methods in 3D. Note however that $N_h = 3$ is likely below the embedding dimension of this system, so we do not claim that we have an exact low-dimensional representation of the system. Moreover, for much more optimized architectures, we refer to \citet{Page2022}. Figure \ref{fig:test_set} shows the performance of the autoencoder on test data. This trajectory was not part of the training data and is hence entirely new to the network. We observe that the network is able to identify the key structures of the trajectory. 

\begin{figure}
    \centering
    \includegraphics[width = \columnwidth]{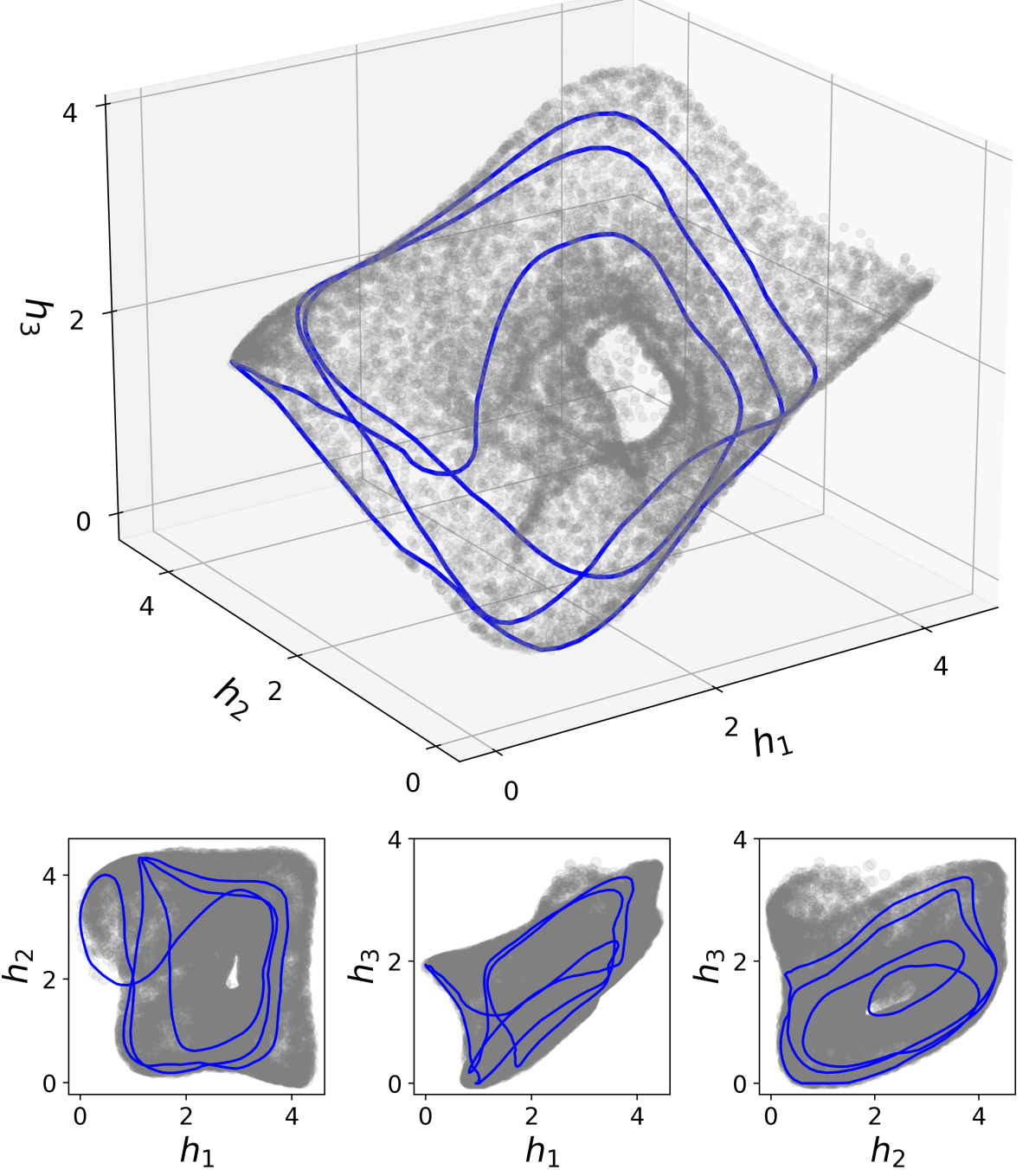}
    \caption{The latent KSE attractor at parameter value ${L = 39}$. The grey dots are random data points from the test and training set. The blue line is a periodic orbit with period $T = 85.536$, plotted to emphasize the shape of the attractor. Top: 3D plot. Bottom: 2D projections.}
    \label{fig:L39latent_attractor}
\end{figure}

Figure \ref{fig:L39latent_attractor} visualises the attractor in the autoencoder's latent space together with a UPO in 3D (top) and in 2D projections (bottom). The points plotted in the figure are part of the data-sets used to train and test the autoencoder. 

\subsubsection{Periodic orbit searches}
We generate multiple loops for different ranges of periods. We use all 3 latent POD modes to generate the loops as all 3 eigenvalues $\gamma_1, \gamma_2, \gamma_3$ are non-zero. As defined in section \ref{sec:algorithm}, we set the number of sine/cosine modes $M$ to be equal to the targeted number of intersections $p$ with an adequate Poincaré section. \citet{Lasagna2017} uses $\hat{u}_1 = 0, (\hat{u}_1)_t > 0$ (where $\hat{u}_1$ is the first component of the Fourier transform of $\boldsymbol{u}$), in which case the dynamics appear to have a return time (the time between two consecutive intersections with the Poincaré section) of approximately 25 time units. Therefore, when we target short orbits with $p = 1$, we let $M = 1$ and choose guess period $T = 25$. For orbits with $p = 2$, we let $M = 2$ to introduce a `twist' in the loop and set the guess period to $T = 50$. For $p = 3$, we introduce two twists by choosing $M = 3$, and so on. For longer UPO searches $(M\geq 3)$, we pick a range of periods between $25M$ and $25(M+1)$. Table V in Appendix \ref{sec:app_tables} shows nicely how the number of Poincaré intersections of the final UPOs scales with $M$.

\begin{figure*}
    \centering
    \includegraphics[width = 1.95\columnwidth]{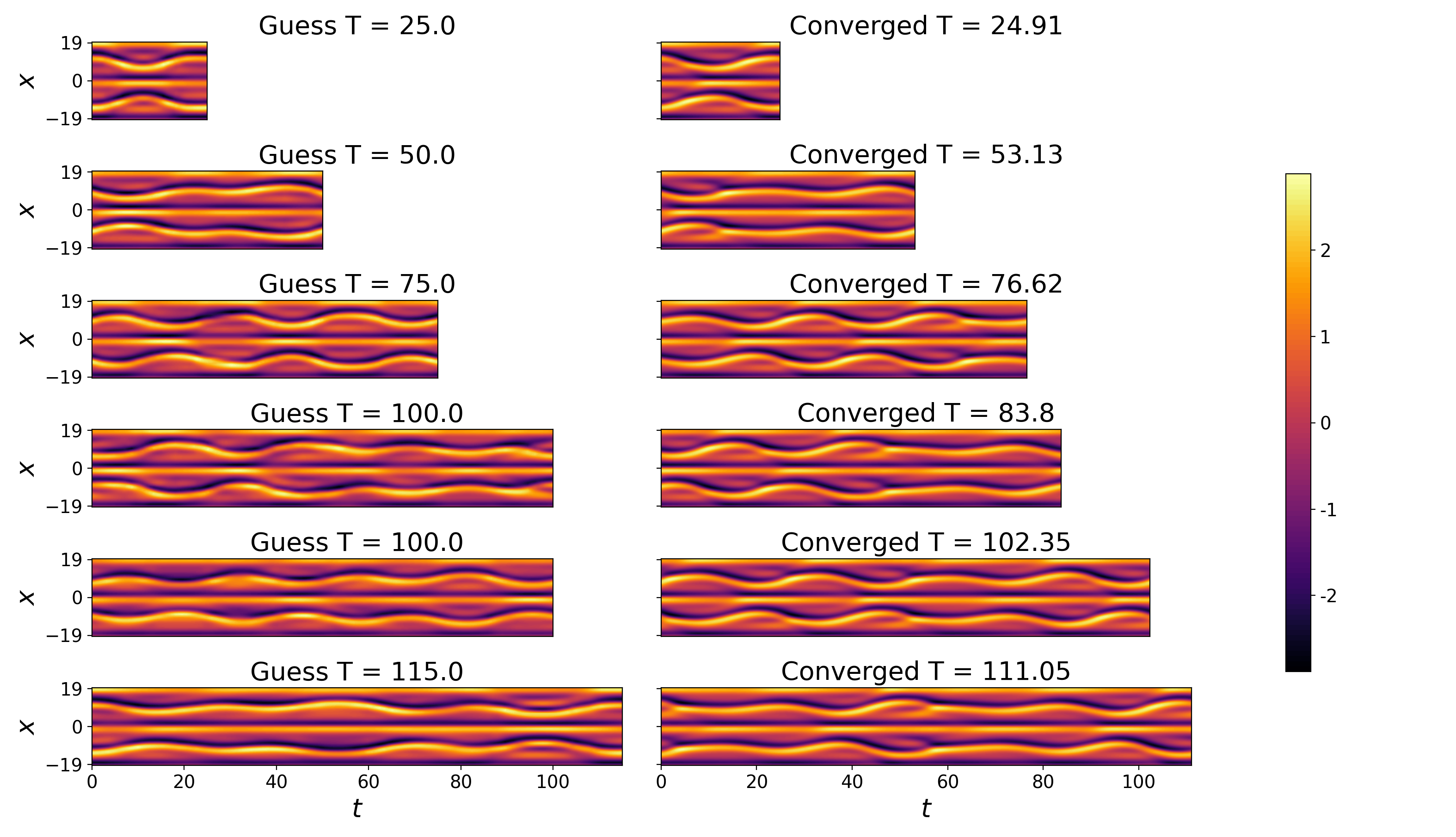}
    \caption{Examples in physical space of loop guesses (left column) that converged to periodic orbits (right column) for various periods. The guesses share a similar sequence of shapes with the converged UPOs.}
    \label{fig:guesses_convs_phys_examples}
\end{figure*}

Figure \ref{fig:guesses_convs_phys_examples} compares decoded guess loops (of various lengths) to the UPOs and periods they converged to. We note that the guesses are realistic as they look like they could be trajectories of the KSE. This is emphasized when comparing the guesses to the final orbit they converge to: the guesses have similar sequences of patterns as the UPOs, and thus they look alike. In general, we find that many of our loops converge to periodic orbits, confirming that they are good initial guesses for loop convergence algorithms. The guesses look realistic and are already close to the UPOs that they eventually converge to. The detailed outputs of the runs for $M = 1,2$ and 3 are given in the tables \ref{tab:25_sweep}-\ref{tab:75_sweep}, where we generated 200, 500 and 700 loops respectively. We also indicate the number of times we converge to fixed points and when the algorithm does not converge. The latter happens either if we stopped the convergence algorithm too early or when the minimization of $J$ gets caught in a local minimum, where $J > 0$ and $\nabla J =0$, instead of converging to a global minimum with $J = 0$. In cases where we converge to multiples of a short orbit (for example twice the 25.37 UPO) in higher-period runs, we cut the orbit into its shortest periodic component and re-converge. We conduct such searches for orbits with Poincaré intersections up to $p = 4$. Every UPO that we find is verified to still exist at temporal resolution $N_t = 256$. We note that for short orbits, we converge very often, with around 70\% or 76\% of guesses converging for targeted Poincaré intersections $p = 1$ and $p = 2$ respectively. As orbits get longer, the success rate naturally drops. While for $p = 3$, around 40\% of the guesses still converge, when we target $p = 4$ this falls to around one quarter. These drops are to be expected as longer UPOs are generally harder to converge to, however we note that when we do latent gluing in section \ref{sec:gluing_app}, the success rate for guesses with large periods shoots up significantly. A detailed overview of all runs is given in table V in Appendix \ref{sec:app_tables}.

\begin{table}[htpb]       
  \centering
  \begin{tabular}{|c|ccc|}
    \hline
    Type          & Count & Percentage & $p$ \\
    \hline\hline
    24.91      & 35    & 17.5       & 1        \\
    25.37      & 104   & 52.0       & 1        \\
    \hline                                        
    No convergence  & 6   & 3.0       &         \\
    Fixed points  & 55    & 27.5       &         \\
    \hline
  \end{tabular}
  \caption{Results for 200 loops with guessed period 25. Left column: type of the result: UPO, no convergence, or fixed point. If converged to a UPO, the period is indicated. No convergence means that the optimization process got stuck in a local minimum, or requires more time to converge. Other columns: number of times (and percentage) the type of result appears. For periodic orbits, the number of intersections $p$ with the Poincaré section is also indicated.}
  \label{tab:25_sweep}
  
  \begin{tabular}{|c|ccc|}
    \hline
    Type          & Count & Percentage & $p$ \\
    \hline\hline
    24.91      & 35    & 7.0        & 1        \\
    25.37      & 20   & 4.0       & 1        \\
    50.37      & 78    & 15.6        & 2        \\
    52.04      & 91    & 18.2        & 2        \\
    53.13      & 157    & 31.4        & 2        \\
    57.23      & 1     & 0.2        & 2        \\
    \hline                                        
    No convergence  & 48   & 9.6       &         \\
    Fixed points  & 70    & 14.0        &         \\
    \hline
  \end{tabular}
  \caption{Results for 500 loops with guessed period 50. Periodic orbits with $p = 1$ reappear when the loop converges to a double periodic orbit. See caption of table \ref{tab:25_sweep} for explanation of terms.}
  \label{tab:50_sweep}
 
  \begin{tabular}{|c|ccc|}
    \hline
    Type          & Count & Percentage & $p$ \\
    \hline\hline
    24.91 & 15 & 2.1 & 1 \\
    25.37 & 4 & 0.6 & 1 \\
    50.37 & 3 & 0.4 & 2 \\
    52.04 & 2 & 0.3 & 2 \\
    53.13 & 3 & 0.4 & 2 \\
    57.23 & 23 & 3.3 & 2 \\
    57.63 & 20 & 2.9 & 2 \\
    75.28 & 35 & 5.0 & 3 \\
    75.72 & 18 & 2.6 & 3 \\
    75.94 & 16 & 2.3 & 3 \\
    76.62 & 40 & 5.7 & 3 \\
    76.85 & 30 & 4.3 & 3 \\
    76.95 & 38 & 5.4 & 3 \\
    77.37 & 16 & 2.3 & 3 \\
    85.54 & 1 & 0.1 & 3 \\
    \hline           
    No convergence & 385 & 55.0 &  \\                             
    Fixed points & 51 & 7.3 &  \\
    \hline
  \end{tabular}
  \caption{Results for 700 loops with guessed periods 75 and 80 (350 loops each). Periodic orbits with $p = 1$ reappear when the loop converges to a triple periodic orbit. See caption of table \ref{tab:25_sweep} for explanation of terms.}
  \label{tab:75_sweep}
\end{table}

\subsubsection{Latent gluing}
\label{sec:gluing_app}
\begin{figure}
    \centering
    \includegraphics[width = \columnwidth]{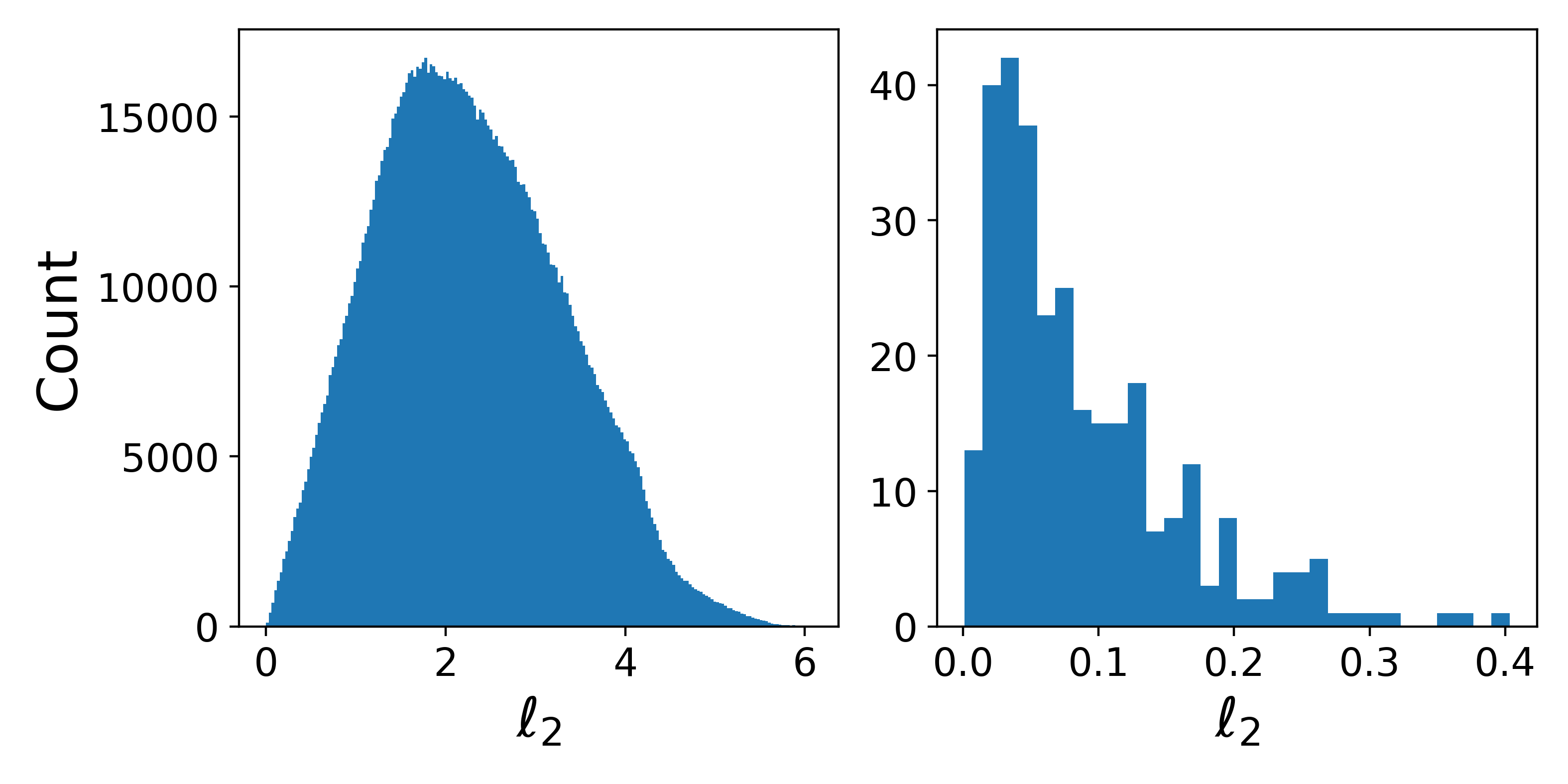}
    \caption{Left: distribution of random $\ell_2$ distances of a long time-series in latent space. Right: distribution of $\ell_2$ distances between points of closest passage between UPOs with periods $T<100$.}
    \label{fig:latent_l2_distribution}
\end{figure}

We create multiple new guesses using the methodology laid out in section \ref{sec:gluing} by gluing the orbits $\mathcal{P}_i$ found above. We limit ourselves to gluing orbits with period $T_i < 100$ for computational efficiency reasons. We run through all possible combinations of gluing two orbits together. Every $\mathcal{P}_i$ has a symmetric counter-part $\mathcal{P}_i^s$ obtained by the shift $x\mapsto x + L/2$. Thus, we glue $\mathcal{P}_i$ with both $\mathcal{P}_j$ and $\mathcal{P}_j^s$. Since we find 18 orbits with $T<100$, this gives a total of 306 combinations. Figure \ref{fig:latent_l2_distribution} compares the distribution of random distances of a long time-series in latent space to the distribution of distances of closest passage $\ell_2$ between two UPOs in latent space and confirms that the points of gluing of two UPOs are indeed close together.

\begin{figure}
    \centering
    \includegraphics[width = \columnwidth]{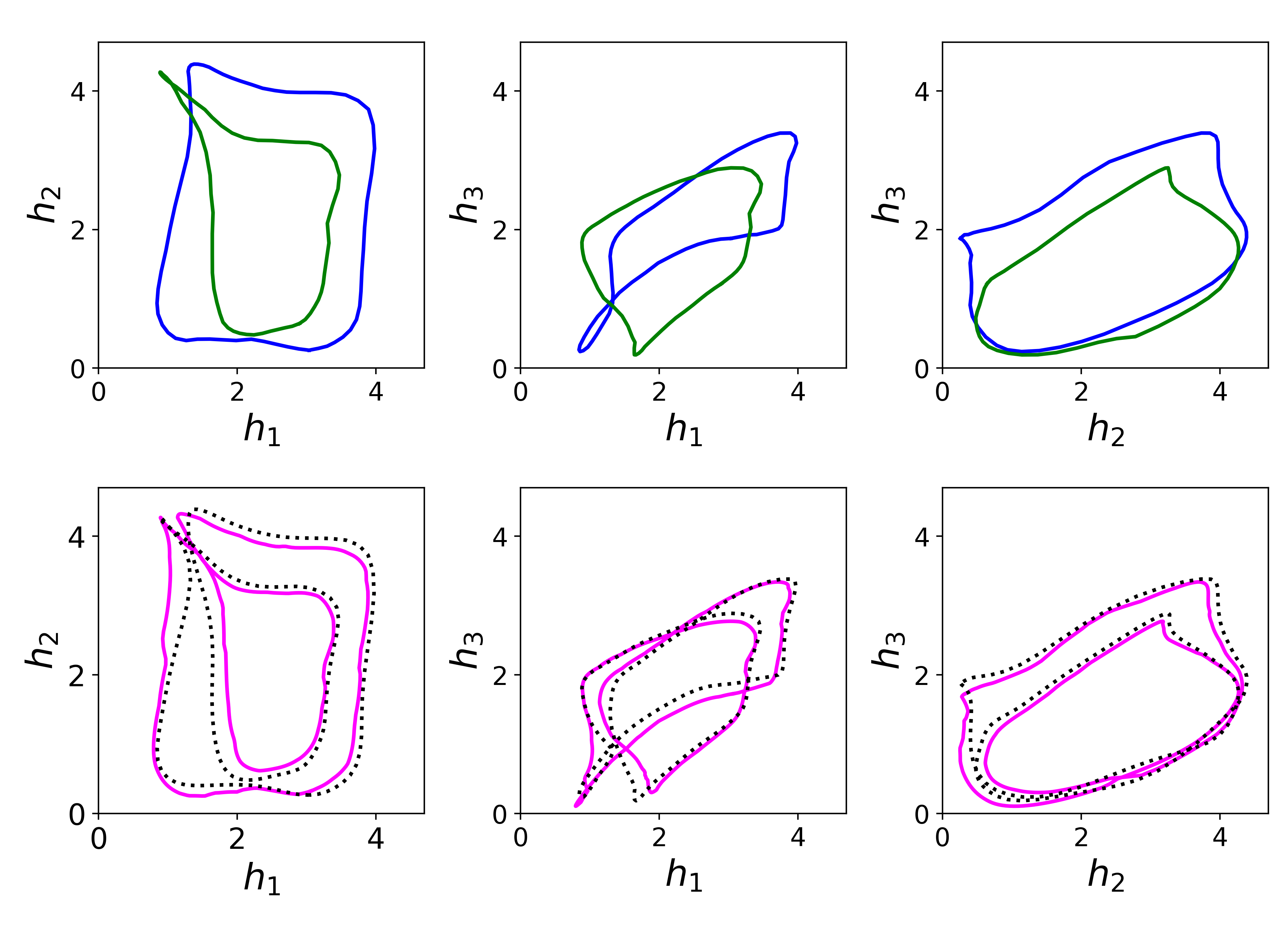}
    \caption{2D projections in latent space of an example of gluing two short UPOs together ($T_1 = 24.908$ and $T_2 = 25.371$). Top: the two orbits (blue and green) are glued together at the points of closest passage. Bottom: The resulting loop is smoothed and serves as a new guess (black dotted), which is then converged to a UPO with period $T = 50.368$ (magenta). The new long UPO resulting from the gluing clearly shadows the two short UPOs.}
    \label{fig:glued_short_latent_2d}
\end{figure}

\begin{figure}
    \centering
    \includegraphics[width = \columnwidth]{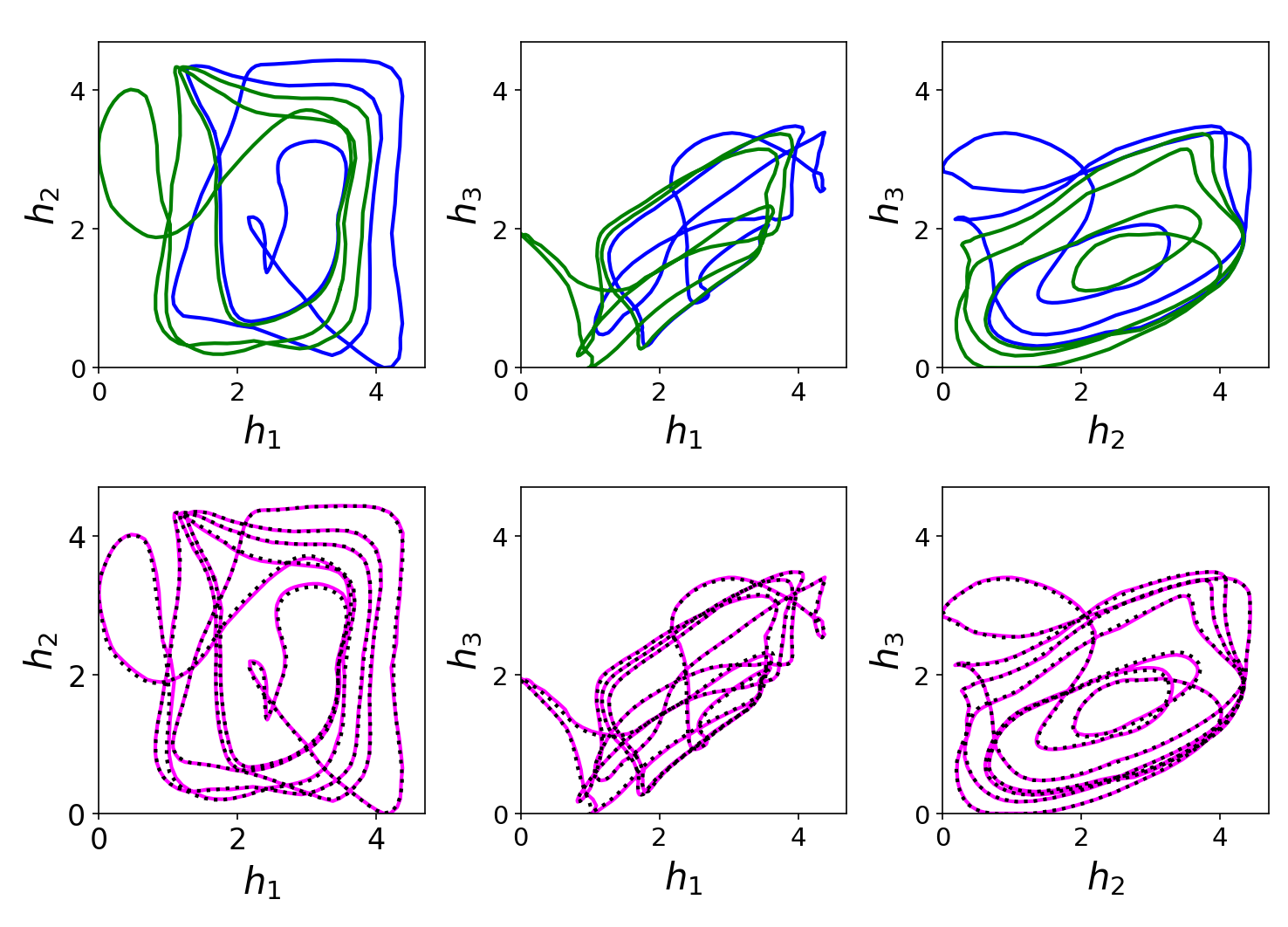}
    \caption{2D projections in latent space of two longer UPOs glued together ($T_1 = 83.804$ and $T_2 = 85.559$). Top: the two individual orbits (blue and green) are glued together at the points of closest passage. Bottom: The resulting loop is smoothed and serves as a new guess (black dotted), which is then converged to a UPO with period $T = 169.467$ (magenta). The new long UPO resulting from the gluing clearly shadows the two short UPOs.}
    \label{fig:glued_long_latent_2d}
\end{figure}

\begin{figure}
    \centering
    \includegraphics[width = \columnwidth]{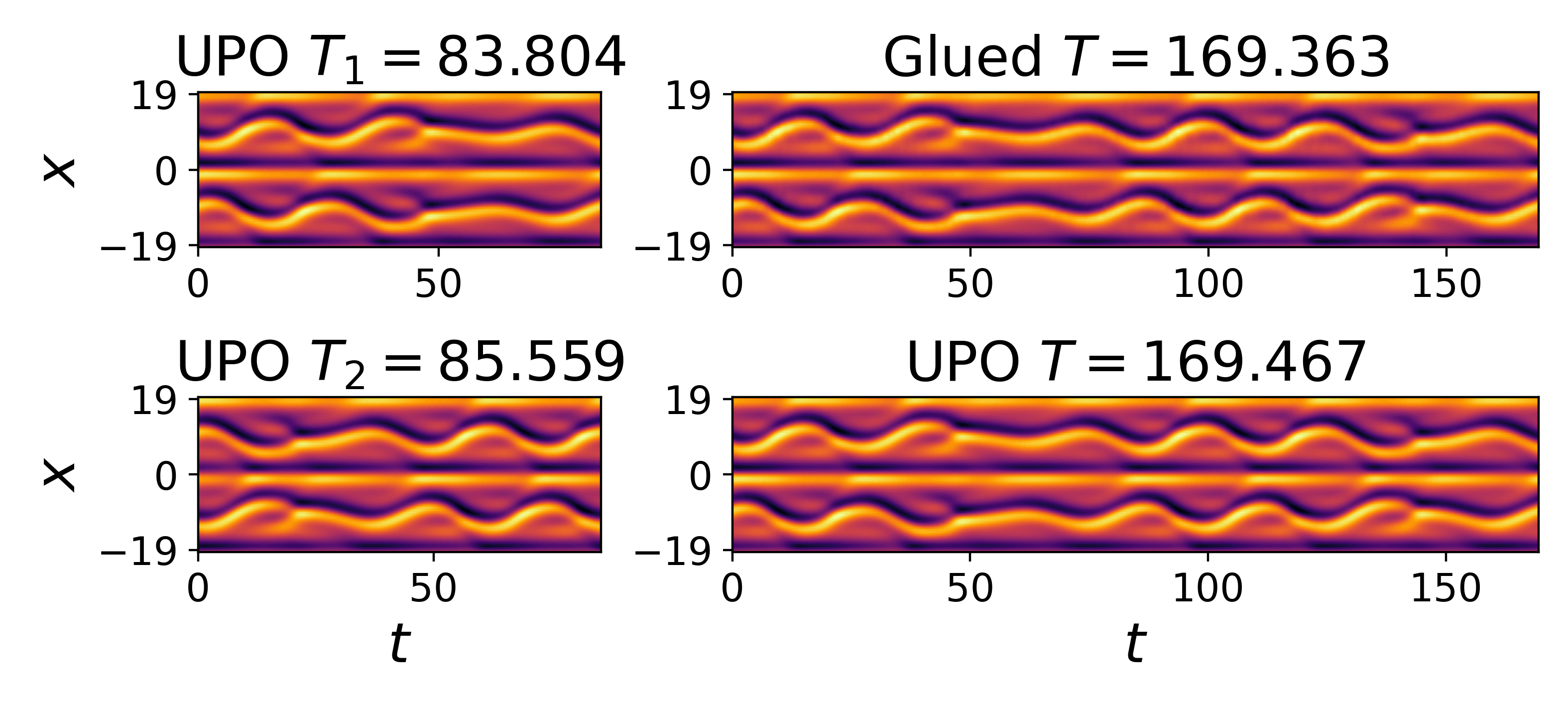}
    \caption{Physical plot of two longer UPOs (left column) with periods $T_1 = 83.804$ and $T_2 = 85.559$ glued together (right column, top), which is then converged to a UPO with period $T = 169.467$ (right column, bottom).}
    \label{fig:glued_long_phys}
\end{figure}

\begin{figure}
    \centering
    \includegraphics[width = \columnwidth]{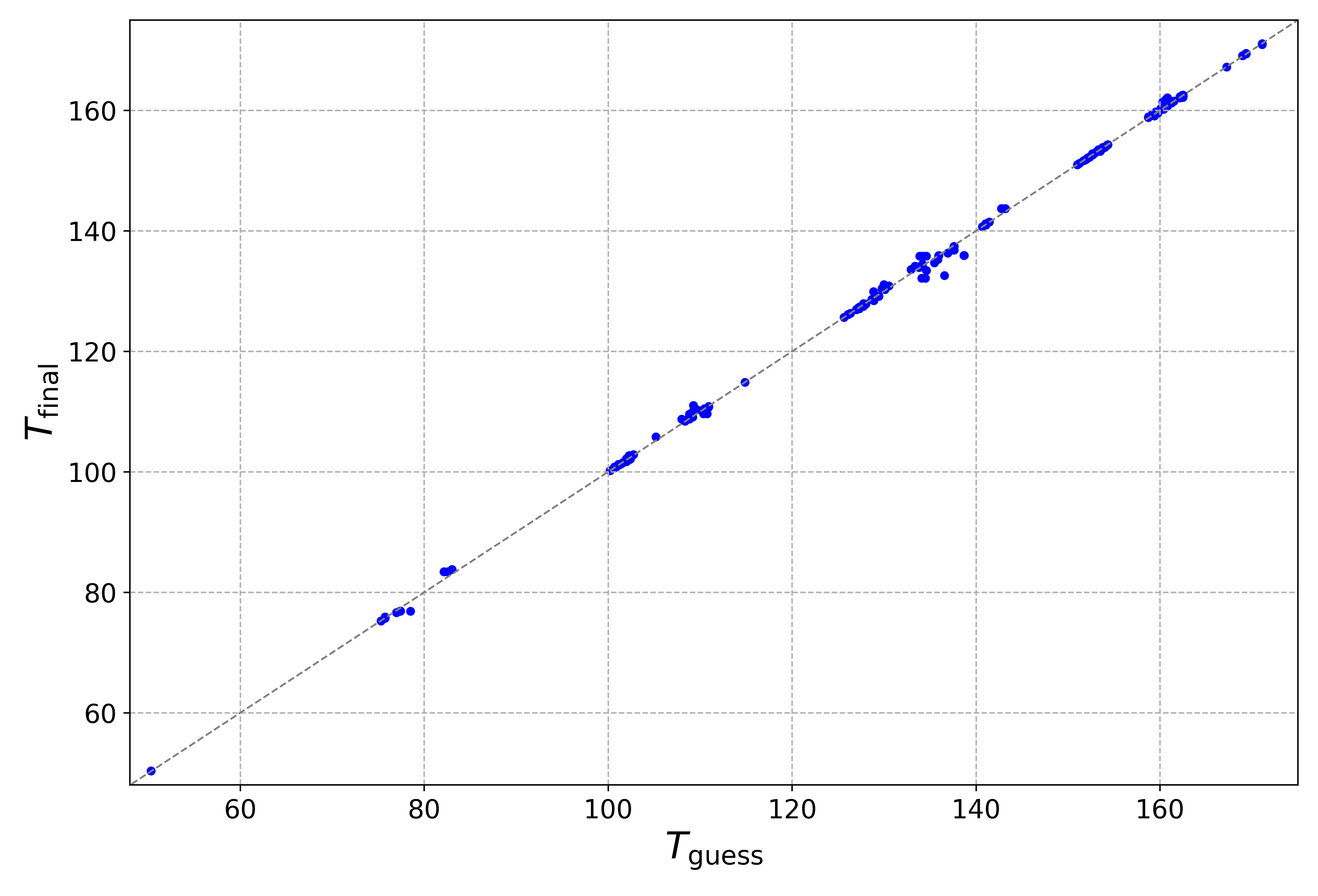}
    \caption{Plot showing the final period the glued guesses converged to against the initial guess period $T_{guess} = T_1 + T_2$.}
    \label{fig:gluing_T_guess_vs_T_final}
\end{figure}

For illustration purposes, figure \ref{fig:glued_short_latent_2d} shows a 2D projection of the gluing process in the latent space between two short orbits with periods $T_1 \approx 24.908$ and $T_2 \approx 25.371$. In this case, the gluing is easy to follow visually, and the glued guess with initial period $T_1 + T_2 \approx 50.279$ converges quickly to a periodic orbit with $T \approx 50.368$. One can easily see that the converged orbit shadows the initial two. Figure \ref{fig:glued_long_latent_2d} shows the same process for two longer initial periodic orbits with periods $T_1 \approx 83.804$ and $T_2 \approx 85.559$. The glued guess with initial period $T_1 + T_2 \approx 169.363$ looks very similar to the converged orbit with period $T \approx 169.467$. This is even more apparent in figure \ref{fig:glued_long_phys}, which shows the decoded, physical plots of these orbits, comparing the glued loop to the final periodic orbit. Again, the converged long orbit appears to shadow the initial two short ones. Figure \ref{fig:gluing_T_guess_vs_T_final} confirms that for the glued guesses that converged, the final period is approximately equal to the sum of the initial periods.

Out of the 306 total guesses, 227 converge to UPOs, 160 of which are distinct, and 79 do not converge (either the optimization of $J$ gets stuck in local minima or we need to run the convergence for longer). The largest period found this way is $T \approx 171.096$ and we find a general success rate of 74.2\%. Impressive is also the success rate for long UPO guesses: loops with guessed periods $T_1 + T_2 > 100$ converge in $210 / 284 \approx 73.9\%$ of cases, compared to the approximately 25\% from purely random guesses (with $M = 4$) observed in the previous section. We note that some UPOs struggle with being glued to other orbits, such as those with periods 53.135, 57.227, and 57.627 (see table VI in Appendix \ref{sec:app_tables}). 

\begin{figure}
    \centering
    \includegraphics[width = \columnwidth]{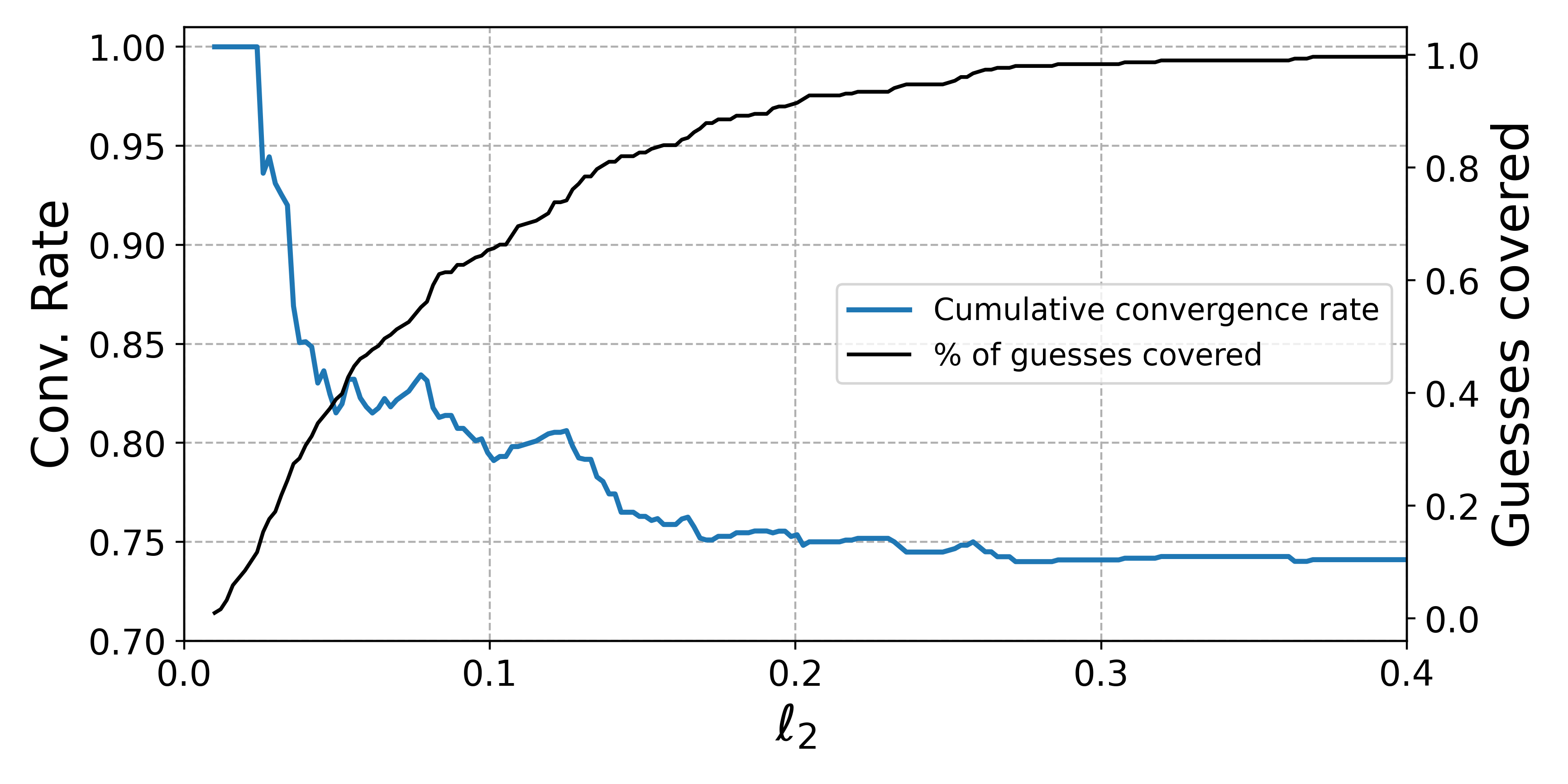}
    \caption{Left axis: Cumulative convergence rate of glued guesses with distance of closest passage in the latent space less than $\ell_2$ (blue). Right axis: percentage of guesses with distance of closest passage in the latent space less than $\ell_2$ (black).}
    \label{fig:gluing_conv_rate}
\end{figure}

Since we expect this hierarchy of periodic orbits, where long orbits shadow shorter ones, we indeed also expect the glued guesses to perform much better than the random ones. Doing the same gluing process directly in physical space yields a similar success-rate of 77.8\%. The key take-away is that the expected hierarchy of UPOs appears to be present in this PDE system, and also in the latent attractor. Moreover, when the $\ell_2$ distance between the two points of closest passage of two UPOs in latent space is small, then the convergence rate is larger. We obtain a convergence rate of 82.2\% for guesses with $\ell_2 < 0.07$. These cover over 50\% of the guesses we attempted. Guesses with $\ell_2 \geq 0.07$ converge in only 65.8\% of cases, showing that this $\ell_2$ is a good indicator of whether orbits are gluable. Figure \ref{fig:gluing_conv_rate} shows the cumulative convergence rate against $\ell_2$.

Out of the total 153 glued symmetry pairs, in 14 cases neither glued guess converges. In 51 cases, we find that while one glued guess does not converge the other converges to a periodic orbit. We also observe that while 39 pairs converge to the same orbit, 49 pairs converge to two distinct ones. For the full output details, see table \ref{tab:gluing_summary} in the appendix.

In summary, in this section we applied the methods described in section \ref{sec:methods} to the KSE in the case of low-dimensional chaos. First, we generated guesses for UPOs by sampling random closed curves in the latent space that on average match the latent flow statistics up to second moments. By varying the number of sine/cosine modes in the linear combination of POD modes, namely the parameter $M$, we observed that the resulting UPOs usually had $p = M$ intersections with the Poincaré section. This allows us to directly target longer UPOs by increasing $M$. Next, we glued UPOs together in the latent space at their closest points of passage. The resulting new guesses had very high convergence rates, indicating that the hierarchy of UPOs observed in ODEs is also present here. This also gives a method to search for longer UPOs. In the next section, we will apply these methods to the hyperchaotic case with $L = 100$.

\subsection{Hyperchaos $L = 100$}
\begin{figure}
    \centering
    \includegraphics[width = \columnwidth]{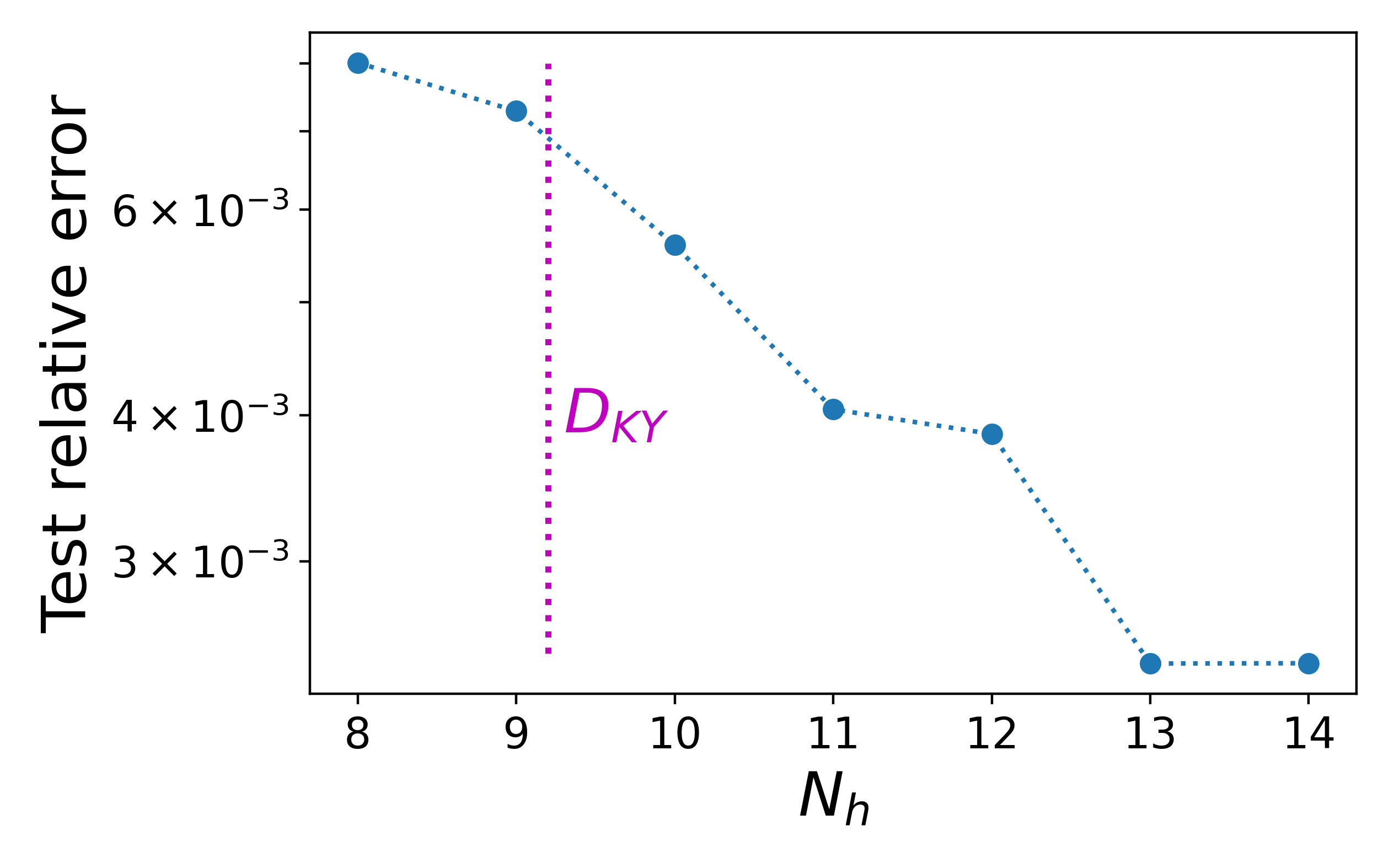}
    \caption{The best test losses of the 20 neural networks trained for each latent dimension $N_h = 8$ to $N_h = 14$ (${D_{KY}\approx 9.2}$ indicated in magenta). For this system we end up choosing $N_h = 11$, however 2 eigenvalues of the latent POD modes are approximately 0, giving 9 effective latent dimensions.}
    \label{fig:L100_network_losses}
\end{figure}

\begin{figure}
    \centering
    \includegraphics[width = \columnwidth]{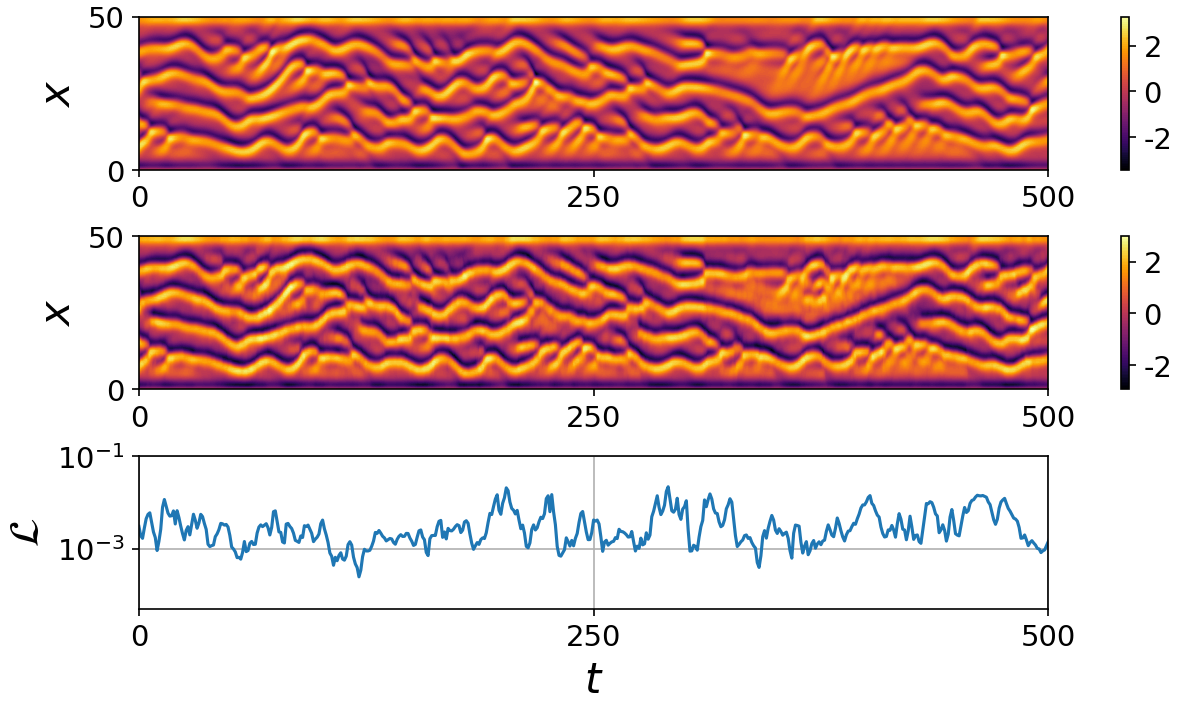}
    \caption{Top: Physical trajectory of the KSE in the test set. This data was not used in training. Middle: Output when applying the autoencoder with latent dimension $N_h = 11$ to the trajectory. Bottom: Loss $\mathcal{L}$ over time.}
    \label{fig:L100_test_set}
\end{figure}

Taking the Kaplan-Yorke dimension $D_{KY} \approx 9.2$ \cite{Edson2019} of the system as guidance, we train 20 autoencoders for each of the latent dimensions $N_h = 8, ..., 14$. Figure \ref{fig:L100_network_losses} shows the best test losses for each value of $N_h$. We note that the losses decrease steadily with latent dimension. By considering the test losses, as well as the $D_{KY}$, we decided to continue with the more parsimonious $N_h = 11$. Figure \ref{fig:L100_test_set} shows the performance of the autoencoder applied to a physical trajectory part of the test set. Although there are some small visual differences between the input and output for the hyperchaotic system, the autoencoder is able to reconstruct the general shapes and structures of the system, and thus shows satisfactory results. Moreover, the latent POD decomposition only has 9 non-zero eigenvalues, resulting in an effective 9 latent dimensions that we use to define our guesses.

\subsubsection{Guesses for the period}
\begin{figure}
    \centering
    \includegraphics[width = \columnwidth]{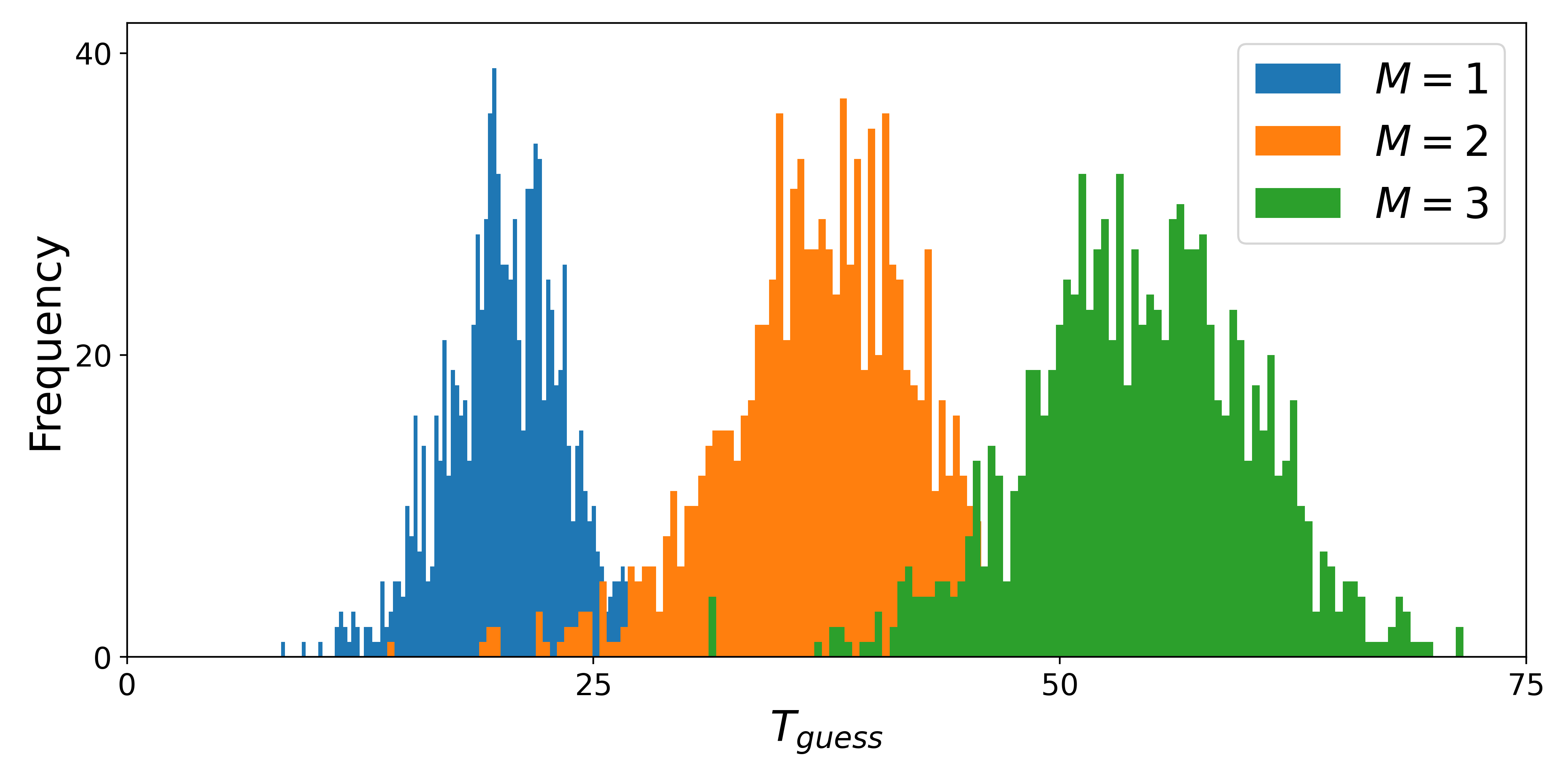}
    \caption{Distribution of $T_{guess}$ calculated for 1000 loops each  at different values of $M$.}
    \label{fig:T_guess_distributions}
\end{figure}

In the $L = 39$ case, we related the period guesses to the system's approximate average return time of the Poincaré section. This becomes less straight-forward in higher-dimensional systems, such as the $L = 100$ case, as a Poincaré section of an $n$-dimensional system should be an $(n-1)$-dimensional subspace. Defining a guess period in this case is non-trivial, however we require one to be able to apply the convergence algorithm from \citet{Azimi2022}.
For a UPO of the system, it is possible to integrate around the UPO and obtain the time taken to traverse it. While a guess for a UPO is not a solution of the KSE, we do have access to $\partial_t u$ at each point of the loop guess by evaluating the right-hand side of the KSE. Therefore, we decided to estimate the period guess of the loop by treating it as if it was a solution of the system and integrating around the loop the time taken to traverse it. More concretely, for a discretized loop time-series $\{\boldsymbol{u}_t\}_{t = 1}^{N_t}$ with $N_t$ time-steps, the guess for the period is

\begin{equation}
    \label{eqn:Tguess}
    T_{guess} = \frac{\sum_{i = 1}^{N_t} || \Delta \boldsymbol{u}_i||}{\frac{1}{N_t}\sum_{i = 1}^{N_t} || \partial_t \boldsymbol{u}_i||}
\end{equation}

where we assume that the time-gap $dt$ between time-steps is constant. The numerator can be interpreted as the length of the loop, and the denominator as the average velocity around the loop. Figure \ref{fig:T_guess_distributions} shows the distribution of $T_{guess}$ for different values of $M$. Similar to the $L = 39$ system, we can aim for longer or shorter UPOs $T_{guess}$ by introducing more "twists" in our loop i.e. by changing $M$.

Note that it would also have been a reasonable choice to use $\partial_t \boldsymbol{u}_i \cdot \hat{\boldsymbol{t}}$ in the denominator of equation \ref{eqn:Tguess}, where $\hat{\boldsymbol{t}}$ is the normalized loop tangent at $\boldsymbol{u}_i$, instead of $||\partial_t \boldsymbol{u}_i||$. The sign of this dot product would tell us whether the guess loop is going in the correct direction at $\boldsymbol{u}_i$. However, since random vectors in high-dimensional spaces are likely to be orthogonal \cite{Goldstein2018PhaseMax:Pursuit}, this results in $T_{guess}$ blowing up and over-estimating the period.

\subsubsection{Periodic orbit searches}
For the initial test searches that we conducted, we observed a sharp increase in $T$ during the adjoint looping procedure related to the increase in parameter $L$. To counteract this and to better target the guess period, we reduce the weight of the period gradient and gradually increase it again as the loss approaches 0. Note that this does not affect the monotonic decrease in $J$: \citet{Azimi2022} choose the adjoint operator $\mathcal{L}^\dagger = -\boldsymbol{G} = -(\boldsymbol{g}_1, g_2)$ so that $\partial_\tau J < 0$, where $\tau$ is the fictitious time. However setting $\boldsymbol{G}' = (\boldsymbol{g}_1, \mu g_2)$ with $\mu > 0$ still results in $\partial_\tau J < 0$. Since we reasonably trust $T_{guess}$, we set $\mu = 0.01$ at the beginning to initially modify the loop itself, and then gradually increase $\mu$ to 1. Moreover, we looked at different values for $\alpha_m = [(m+1)/(M+1)]^\beta $ by varying the exponent $\beta$ that distributes the weighting between smaller and higher modes. We tried $\beta = \frac{1}{2}, 1, 2$ and $M$, and found the best performance again with $\beta = 1$.

\begin{figure*}
    \centering
    \includegraphics[width = 1.5\columnwidth]{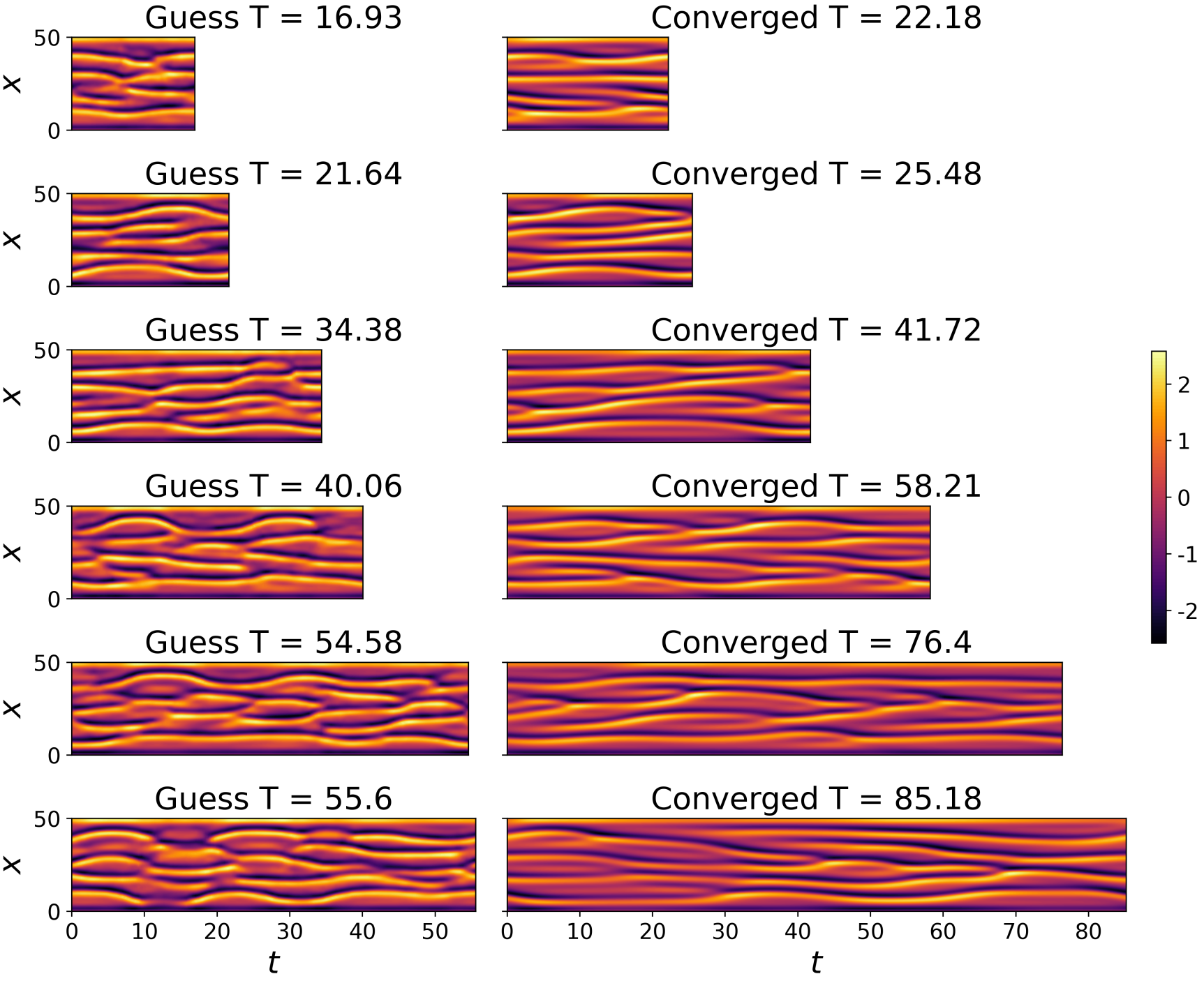}
    \caption{Examples in physical space of guess loops (left column) that converged to periodic orbits (right column) for various periods. The guesses share a similar sequence of shapes with the converged UPOs.}
    \label{fig:L100_guesses_convs_phys_examples}
\end{figure*}

We conduct searches for UPOs based on guesses generated with $M = 1,2$ and 3. For each value of $M$, we generate 1,000 guesses. A summary of these three searches and their results is given in table \ref{tab:searches_100}. Examples of guesses generated from the autoencoder and the periodic orbits they converge to are shown in figure \ref{fig:L100_guesses_convs_phys_examples}. As for the $L = 39$ case, we observe that the guesses and the periodic orbits that they converge to share similar shapes. A noticeable difference to $L = 39$ case is that during the convergence, the loops undergo a stretching and squeezing process with certain parts of the orbit being traversed faster or slower than for the guess, and the initial period often being an underestimate. This highlights the challenge of correctly parametrizing the initial guess inside the latent space. $T_{guess}$ appears to underestimate the period, which may be due to a bias from our Riemann-sum approach in the definition of $T_{guess}$.

In general, we observe good convergence rates for short orbits, with 153 orbits converged for $M = 1$, of which 130 are distinct. We note that the converged periods create more of a continuous spectrum (rather than multiples of an average return time as in the $L = 39$ system), again reflecting the increased complexity of this system. The success rate naturally drops for $M = 2$ and 3, going to 49 and 11 orbits respectively (all of which are distinct). We attribute this drop to various factors, namely the much increased complexity of the system at $L = 100$, the ad-hoc definition for guesses based on sines and cosines without taking into account the dynamics apart from the moment matching, the difficulty in correctly parametrizing the speed of traversing the guess in the latent space, and how efficient the algorithm that we used is at converging long UPOs. Indeed, since long UPOs require more time-steps for accurate convergence, it is likely that if we let the convergence run for longer that more of our guesses would converge. Nevertheless, the ability to find many short orbits with such ease is a success in itself: the generation of these guesses only requires a one-time up-front cost in training the network. Once this is done, guesses can be generated cheaply and instantaneously.

\begin{table}[htpb]
  \centering
  \begin{tabular}{|c|c|c|c|}      
    \hline 
    $M$ & 1 & 2 & 3 \\ 
    \hline 
    Guesses & 1,000 & 1,000 & 1,000 \\
    Fixed points & 13 & 0 & 0 \\
    No convergence & 834 & 951 & 989 \\
    UPOs & 153 & 49 & 11 \\
    \hline
    \end{tabular}
  \caption{Summary of the main UPO searches at $L = 100$ for $M = 1,2$ and 3. The success rate clearly drops as $M$ increases, which may be due to multiple factors, such as the crudeness of the guess definition or stopping the convergence too early.}
  \label{tab:searches_100}
\end{table}

\begin{figure}
    \centering
    \includegraphics[width = 1\columnwidth]{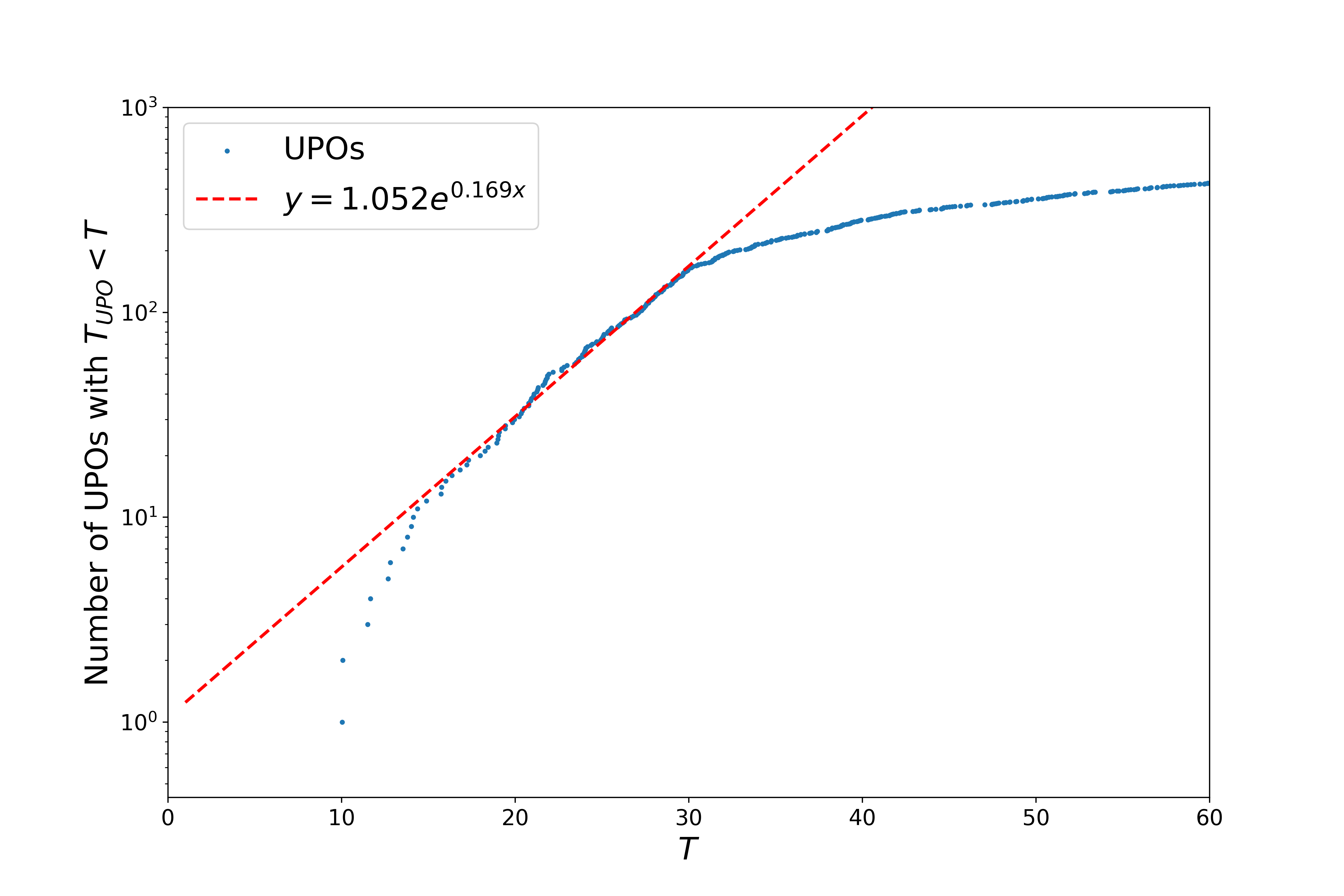}
    \caption{Number of periodic orbits found up to a given period $T$ (blue) as in \citet{Cvitanovic2010a}. We plot a regression of the exponential trend in the number of periodic orbits (red) based on $T\in [20,30]$.}
    \label{fig:period_orbits_exponential}
\end{figure}

From the three main searches we obtained 213 UPOs, 190 of which are distinct, with periods ranging from 10.03 to 110.11. As mentioned earlier, we also conducted other searches to examine different values of $\alpha_m$, different discretization methods, and also initial searches without an adequately weighted period gradient. During these less successful searches, we also found other periodic orbits, giving us a total of 492 distinct orbits, with periods ranging from 9.96 to 110.11. All UPOs that we found were verified to still exist at temporal resolution $N_t = 256$ and spatial resolution $N_x = 256$. Figure \ref{fig:period_orbits_exponential} shows the number of periodic orbits that we have found up to a given period. Based on \citet{Cvitanovic2010a} we expect the number of UPOs to increase exponentially for large periods. On figure \ref{fig:period_orbits_exponential} we plot the exponential trend for orbits with periods $T\in [20,30]$. For $T > 30$ we have not found enough UPOs to continue the trend. For such long UPOs, we expect the structure of the loop state space to be much more complex, implying the existence of many more local minima and thus less successful convergence rates (as also observed in \citet{PageHoley2023}). 

\subsubsection{Latent gluing}
 Since we have a set of 492 UPOs available to us, an exhaustive gluing approach would give us a total of over 241,000 glued guesses. We shall not attempt all of these for computational efficiency reasons. Using the results from section \ref{sec:gluing}, we will restrict ourselves to the following two searches:
 \begin{itemize}
     \item \textbf{Search A}: Guesses where the initial two orbits are close in the latent space, with $\ell_2 < 0.1$, and where $T_1 + T_2 < 125$. This gives a total of 877 guesses.
     \item \textbf{Search B}: A random selection of 1,000 glued guesses among those with $T_1 + T_2 < 125$.
 \end{itemize} 
 
\begin{figure}
    \centering
    \includegraphics[width = \columnwidth]{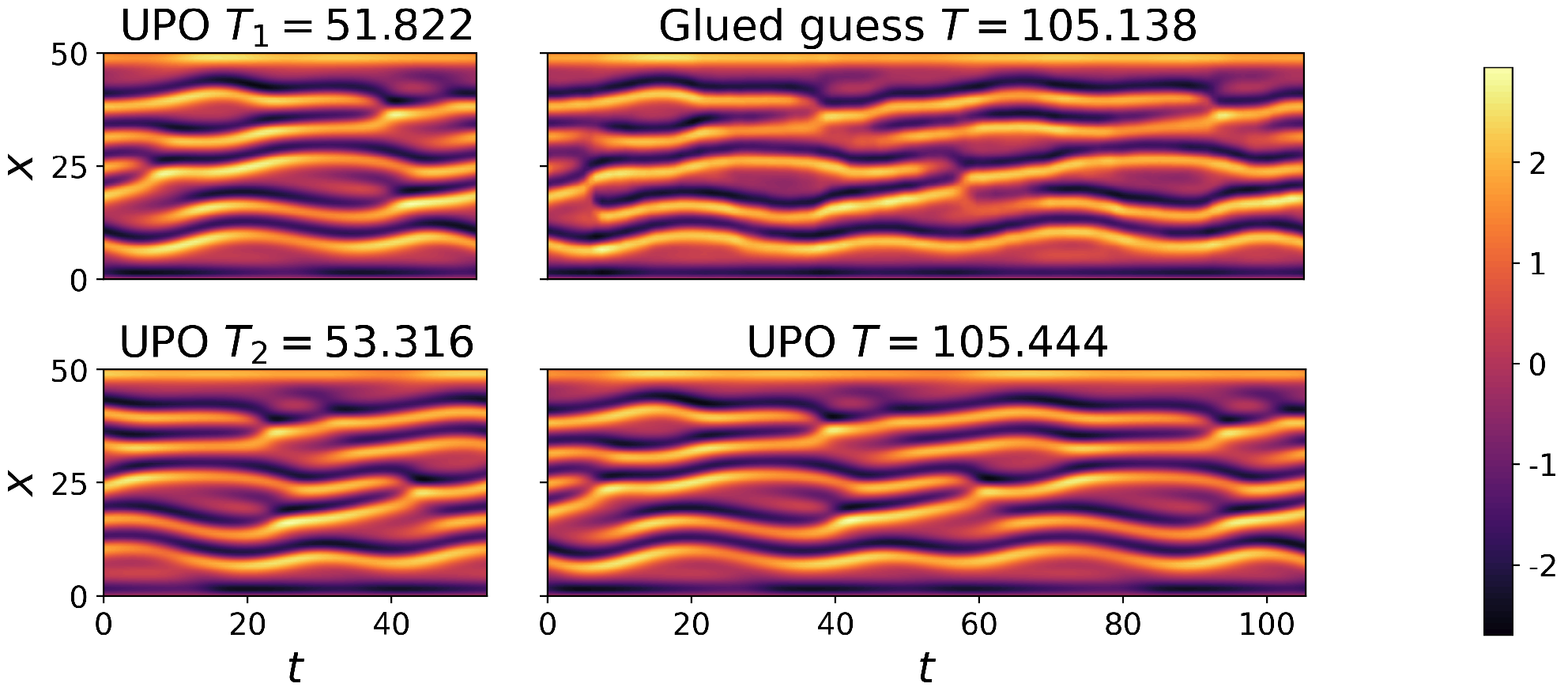}
    \caption{Physical plot of two UPOs (left column) with periods $T_1 = 51.822$ and $T_2 = 53.316$ glued together (right column, top), which is then converged to a UPO with period $T = 105.444$ (right column, bottom).}
    \label{fig:L100glued_long_phys_crop}
\end{figure}

\begin{figure}
    \centering
    \includegraphics[width = \columnwidth]{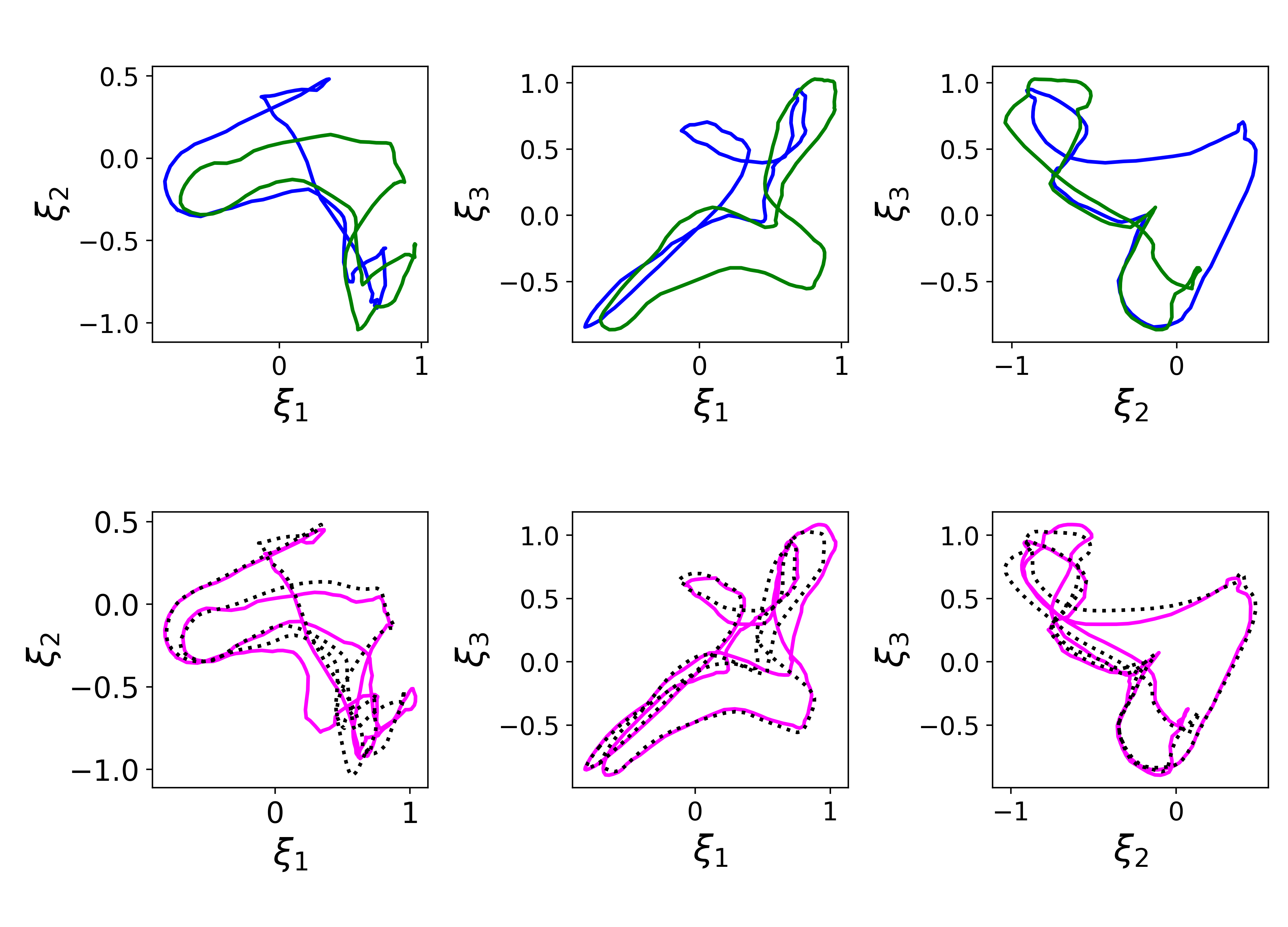}
    \caption{2D projections onto the first three latent POD modes of the gluing process, highlighting how the resulting long UPO shadows the two short UPOs. Top: the two individual orbits ($T_1 = 51.822$ in blue, $T_2 = 53.316$ in green) are glued together at the points of closest passage. Bottom: The resulting loop is smoothed and serves as a new guess (black dotted), which is then converged to a UPO with period $T = 105.444$ (magenta).}
    \label{fig:L100glued_long_latent_2d}
\end{figure}

We restrict ourselves to $T_1 + T_2 < 125$ for computational efficiency again, since the larger the expected period $T$, the higher the time-resolution should be and the longer it takes to compute these orbits. Physical plots of a typical example of a successful gluing between two UPOs are shown in figure \ref{fig:L100glued_long_phys_crop}. Figure \ref{fig:L100glued_long_latent_2d} shows the 2D projections of the initial orbits and the final one onto the first three latent POD modes. We note that the initial UPOs appear embedded within the glued UPO, in the sense that the long glued UPO shadows the shorter ones. 

\begin{figure}
    \centering
    \includegraphics[width = \columnwidth]{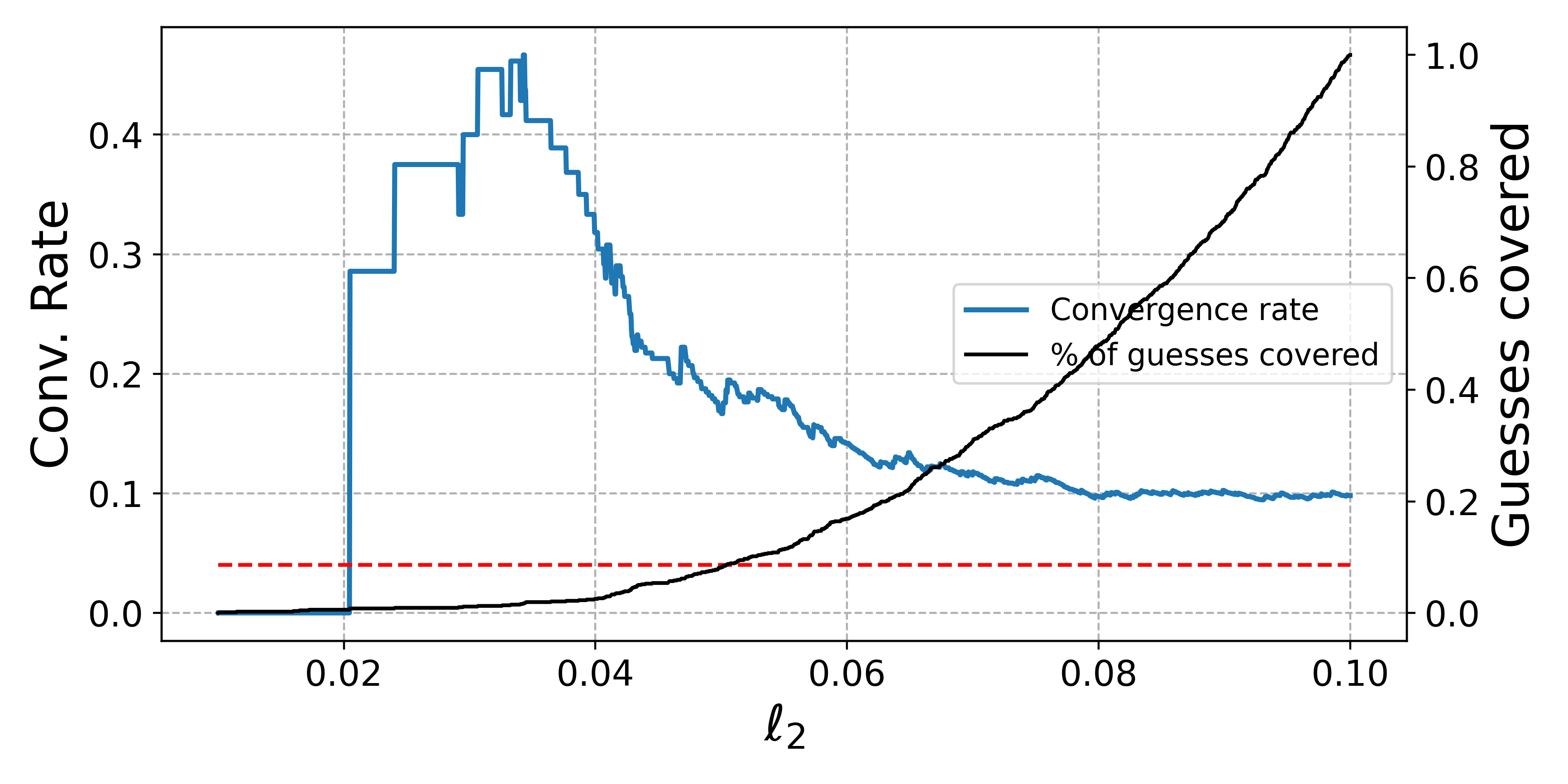}
    \caption{Left axis: Cumulative convergence rate of glued guesses in search A with distance of closest passage in the latent space less than $\ell_2$ (blue). Right axis: percentage of guesses in search A with distance of closest passage in the latent space less than $\ell_2$ (black). Red dashed: Convergence rate of search B. The convergence rate for closer orbits is noticeably larger than for random orbits.}
    \label{fig:L100gluing_conv_rate}
\end{figure}

Search A results in a success rate of approximately 9.8\%, with the cumulative convergence rates shown in figure \ref{fig:L100gluing_conv_rate}. In particular, looking at the 88 closest orbits (the closest 10\%), the convergence rate is 18.2\%. Search B, which consists of 1,000 random glued guesses, only has a success rate of 3.8\%, clearly showing that a small $\ell_2$ increases the likelihood of two orbits being able to be glued together. This indicates that also for the hyperchaotic system there seems to be a hierarchy of UPOs, where long UPOs shadow shorter ones, both in the physical space and in the autoencoder's latent space.

The success rate for $L=100$ is noticeably smaller than for $L = 39$. While this may be partly attributed to the higher complexity of the system, it also indicates that there might be potential for optimizing such gluing procedures. Of course, since glued orbits have a larger expected period, attempting to converge these orbits for longer might also increase the success rate.

\section{Conclusion \& discussion}
\label{sec:conclusion}
In this paper we have introduced a new method for generating initial guesses for periodic orbits by randomly drawing loops in the low-dimensional latent space defined by an autoencoder. The autoencoder's latent dimension was chosen in a process where multiple networks were trained for various latent dimensions. Among these, we decided to work with a latent dimension that shows a satisfactory loss and is in agreement with the Kaplan-Yorke dimension. The loop guesses are constructed by linear combinations of the latent POD modes (that have non-zero eigenvalues) with periodic coefficients drawn from a random distribution that match the latent flow statistics. These loops are then decoded back to the physical space and together with a guess period serve as an initial guess for a loop convergence algorithm. We apply this method to the Kuramoto-Sivashinsky PDE in regimes of low-dimensional chaos and hyperchaos. The decoded loops lie close to the chaotic attractor, look realistic and prove to be good guesses for periodic orbits in loop convergence algorithms, with many of them converging to periodic orbits. This provides an alternative to recurrence methods, where guesses are based on near recurrences in a long DNS, and are thus rare and expensive to generate, while also being biased towards short and less unstable periodic orbits. We note that the derivation for loop guesses based on POD modes holds for both the physical and the latent space. For systems exhibiting low-dimensional chaos, defining guesses based on the physical POD modes is a valid approach and gives acceptable guesses. However, we observed that with guesses purely based on the physical POD modes, the convergence rate appears to be generally lower, and we tend to converge to the same few UPOs. As the system becomes more complicated, it appears necessary to rely on a nonlinear order reduction method, like autoencoders, and define loops in the latent space via the latent POD modes to enforce a larger variation of loop guesses that approximately lie on the latent attractor.

Motivated by a hierarchy of UPOs that is present in ODE systems, such as the Lorenz equations, we explored the method of latent gluing, where we concatenate two periodic orbits based on where they are closest to each other in the latent space. After smoothing this new loop inside the latent space and decoding it back to the physical space, we use this as an initial guess for longer periodic orbits, as we expect the hierarchy of UPOs to carry over from ODEs to PDEs. Many of these guesses converge to UPOs, indicating that this hierarchy does indeed hold in the specific PDE system, with long UPOs shadowing shorter ones. This motivates further research into whether this hierarchy is also present in more complicated spatiotemporally chaotic PDE systems, such as the Navier-Stokes equations. Additionally, the gluing provides a method for generating new, longer UPOs from known shorter ones. Importantly, in both regimes of low-dimensional chaos and hyperchaos, the gluing is much more successful if the distance of closest passage between the two initial UPOs is smaller. This indicates that the hierarchy is also present in the autoencoder's latent space and that the autoencoder is able to capture a coherent low-dimensional representation of the system. For both the chaotic and hyperchaotic systems we conclude that a small distance between the points of closest passage of two periodic orbits in latent space significantly increases the likelihood of whether two UPOs are `gluable'. 

These results are a step forward for using loop convergence methods to find unstable periodic orbits, as we have shown that we can easily and cheaply generate good initial guesses. As expected, there are limitations when moving to long UPOs in hyperchaos. However, this also constitutes the outlook for the future: we believe that optimizing the autoencoder, as well as the guesses themselves so that they are built on knowledge of the dynamics will significantly improve this method. 
Having identified the hierachical organization of UPOs with long orbits shadowing a sequence of shorter ones moreover fuels the hope to generate symbolic encodings of those UPOs not only for simple ODEs but for formally infinite-dimensional PDE systems and thereby - at least approximately - enumerate UPOs which is of key relevance when trying to transfer ergodic theory concepts from chaotic ODEs to dissipative nonlinear chaotic PDEs underlying physically relevant phenomena such as fluid turbulence.

\section{Acknowledgements}
The authors thank Omid Ashtari for his invaluable insights on loop convergence methods as well as his code. This work was supported by the European Research Council (ERC) under the European Union’s Horizon 2020 research and innovation programme (grant no. 865677).

\appendix

\section{Training specifications}\label{sec:app_training}
The networks were created with TensorFlow 2.11.0 and trained on a NVIDIA RTX A4000 GPU. The $L = 39$ networks had inputs of size 31 and the encoder part consisted of three dense layers with 32, 32 and $N_h$ nodes respectively. The decoder has the inverse setup and consists of three dense layers of sizes 32, 32 and 31. Each layer has ReLU activation functions. We trained 20 networks for each $N_h\in\{1,2,3,4,5\}$ for 500 epochs. 
The $L = 100$ networks had inputs of size 84 and the encoder part consisted of three dense layers with 256, 128 and $N_h$ nodes respectively with ReLU activation functions, followed by 4 linear layers with $N_h$ nodes for implicit rank minimization. The decoder consists of three dense layers of sizes 128, 256 and 84, each followed by a ReLU activation function. We trained 20 networks for each $N_h\in\{8,9,10,11,12,13,14\}$ for 1,000 epochs.

For all networks we used the AdamW optimizer and an initial learning rate of $7*10^{-4}$ with exponential decay. Each network also had $l_2$ regularization.

\begin{figure}
    \centering
    \includegraphics[width = 0.7\columnwidth]{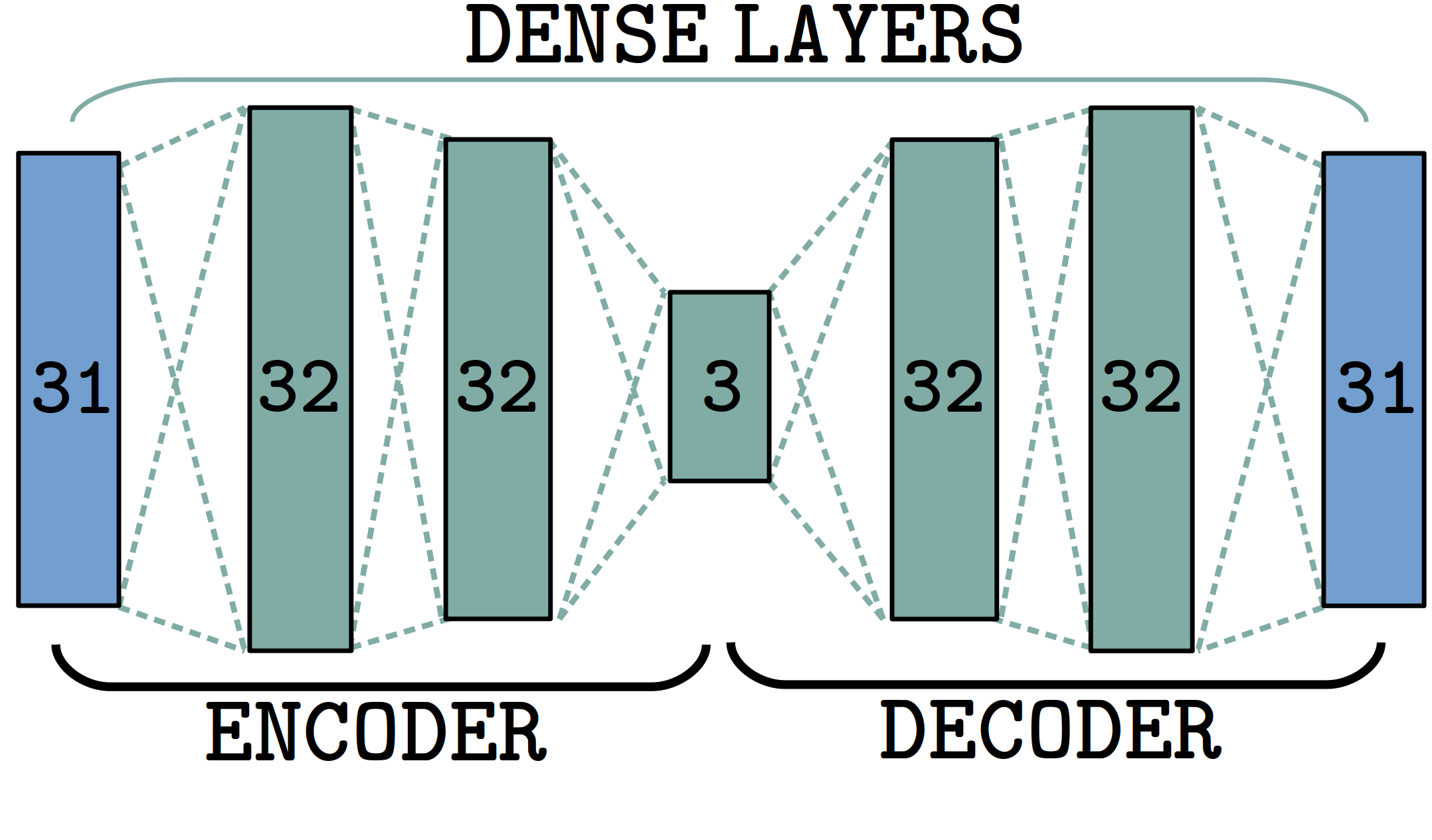}

    \centering
    \includegraphics[width = \columnwidth]{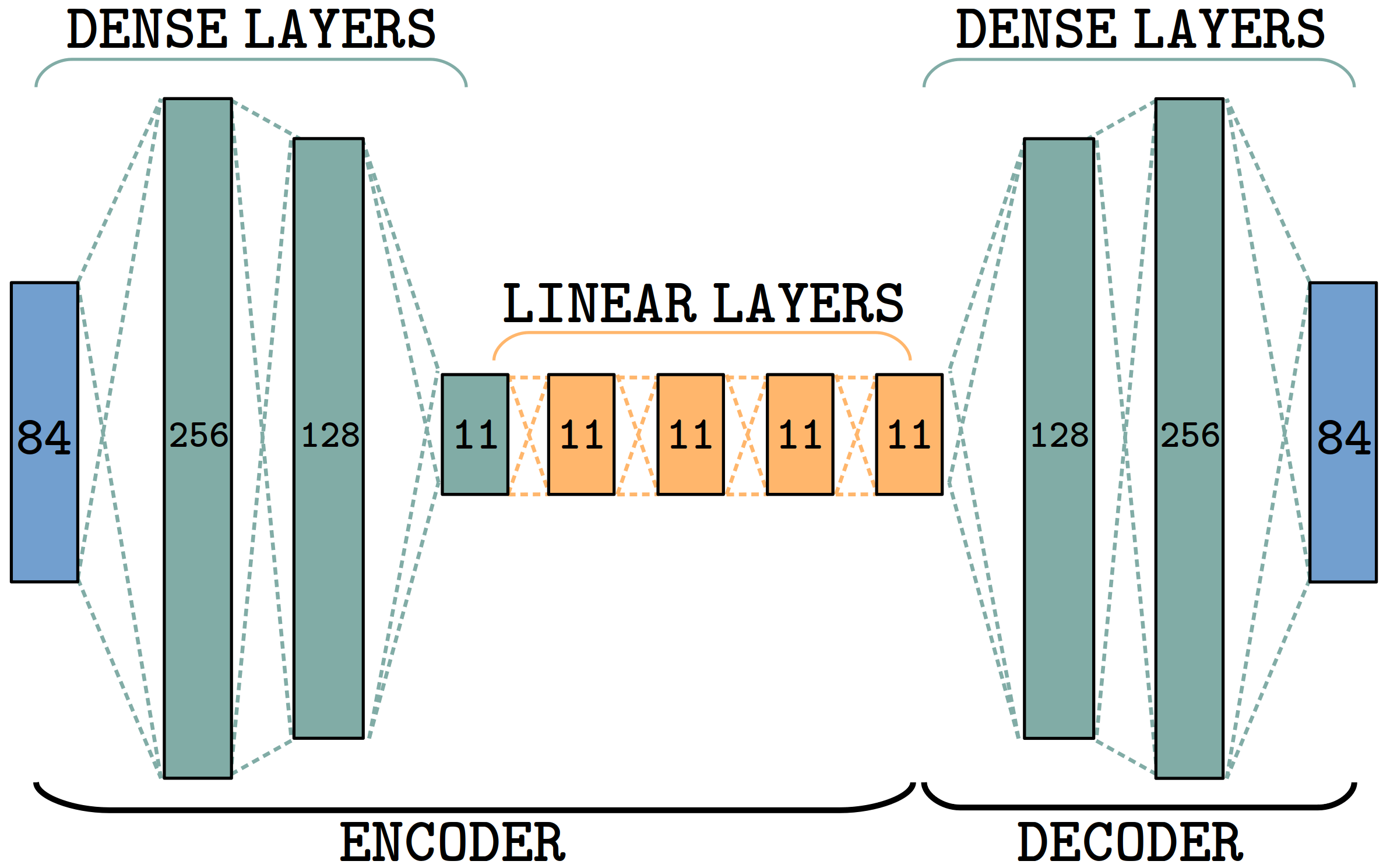}
    \caption{Autoencoder architecture for $L = 39$ (top) and $L = 100$ (bottom, with linear layers for implicit rank minimization \cite{Jing2020AE}). The number in each layer indicates the number of nodes. The dense layers use ReLU activation function.}
    \label{fig:autoencoders}
\end{figure}

\section{Loops based on POD modes}
\subsection{Matching moments}\label{sec:moments}
The $a_k$ are independent as we assume the $X_{m,k}$ to be iid. Since $\mathbb{E}_{X,s}$ is linear and the POD modes form an orthonormal basis, equation \ref{eqn:eq_mean} is satisfied if and only if $\mathbb{E}_{X,s}[a_k] = 0$. Using this in equation \ref{eqn:eq_cov}, we find that

\begin{align*}
    &C^{(\boldsymbol{L})}_{ij} = \mathrm{cov}_{X,s}(L_i, L_j) \\
    &= \mathrm{cov}_{X,s}\Bigg(\mean{u}_i + \sum_{k = 1}^N a_k(\boldsymbol{\phi}_k)_i, \mean{u}_j + \sum_{k = 1}^N a_k(\boldsymbol{\phi}_k)_j\Bigg) \\
    &= \mathrm{cov}_{X,s}\Bigg(\sum_{k = 1}^N a_k(\boldsymbol{\phi}_k)_i, \sum_{k = 1}^N a_k(\boldsymbol{\phi}_k)_j\Bigg) \\
    &= \mathbb{E}_{X,s}\Bigg[ \sum_{p = 1}^N a_p(\boldsymbol{\phi}_p)_i \sum_{q = 1}^N a_q(\boldsymbol{\phi}_q)_j\Bigg] \\
    &= \mathbb{E}_{X,s}\Bigg[ \sum_{p, q = 1}^N a_p a_q(\boldsymbol{\phi}_p)_i (\boldsymbol{\phi}_q)_j\Bigg] \\
    &= \sum_{p, q = 1}^N \mathbb{E}_{X,s}[a_p a_q] (\boldsymbol{\phi}_p)_i (\boldsymbol{\phi}_q)_j \\
    &\implies \boldsymbol{C}^{(\boldsymbol{L})} =  \sum_{p, q = 1}^N \mathrm{cov}_{X,s}(a_p, a_q) \boldsymbol{\phi}_p \boldsymbol{\phi}_q^T
\end{align*}
Since the $a_k$ are independent from each other, then $\mathrm{cov}_{X,s}(a_p, a_q) = 0$ when $p\neq q$ and $\mathrm{cov}_{X,s}(a_p, a_q) = \mathrm{var}_{X,s}(a_p)$ if $p = q$, giving 
\begin{equation}
    C^{(\boldsymbol{L})}_{ij} = \sum_{p = 1}^N \mathrm{var}_{X,s}(a_p) (\boldsymbol{\phi}_p)_i (\boldsymbol{\phi}_p)_j
    \label{eq_CLij}
\end{equation}
Now note that $\boldsymbol{C} = \boldsymbol{V}\boldsymbol{D}\boldsymbol{V}^T$, where $\boldsymbol{V}$ has columns $\boldsymbol{\phi}_1, ..., \boldsymbol{\phi}_N$ and $\boldsymbol{D} = \mathrm{diag}(\lambda_1, ..., \lambda_N)$. Hence
\begin{align*}
    C_{ij} &= \sum_{p,q=1}^N V_{ip} D_{pq} V_{qj}^T \\
    &= \sum_{p=1}^N \lambda_p V_{ip} V_{jp} \\
   \implies C_{ij}  &= \sum_{p=1}^N \lambda_p(\boldsymbol{\phi}_p)_i (\boldsymbol{\phi}_p)_j
\end{align*}
By comparing this expression to equation \ref{eq_CLij}, we can match second moments by setting 
\begin{equation}
    \mathrm{var}_{X,s}(a_k) = \lambda_k
\end{equation}
for $k = 1, ..., N$.

\subsection{Deriving coefficients}\label{sec:coeffs}
We define the coefficients to be a sum of sines and cosines
\begin{equation}
    \label{eqn:sin_cos}
    a_k(s, A_{:,k}, B_{:,k}) = \sum_{m = 0}^M \alpha_{m} [A_{m, k} \cos(ms) - B_{m, k} \sin(ms)]
\end{equation}
where $M$ is the number of sine/cosine modes to be included in the sum, and the coefficients $A_{m, k}, B_{m, k} \sim X$ are iid. The $\alpha_{m}$ are constants that give different weights of choice to higher frequency terms, for example ${\alpha_{m} = (m + 1) / (M + 1)}$. Then
\begin{multline}
    \mathbb{E}_{X,s}[a_k] = \sum_{m = 0}^M \alpha_{m} \{\mathbb{E}_X[A_{m, k}] \langle\cos(ms)\rangle_s \\ - \mathbb{E}_X[B_{m, k}] \langle\sin(ms)\rangle_s\}
\end{multline}
Since the $s$-integrals are 0 for $m>0$, this only requires $\mathbb{E}_X[A_{0,k}] = 0$. To simplify the variance calculation, we will set $\mathbb{E}_X[A_{m,k}] = \mathbb{E}_X[B_{m,k}] = 0$ for all $m$. The variance is given by
\begin{align*}
    &\mathrm{var}_{X,s}(a_k) = \mathbb{E}_{X,s}\bigg[\bigg(\sum_{m = 0}^M \alpha_{m}(A_{m, k} \cos(ms) \\
    & \qquad\qquad\qquad\qquad - B_{m, k} \sin(ms) )\bigg)^2\bigg] \\
    & = \sum_{m,n = 0}^M \alpha_{m}\alpha_{n}\bigg( \mathbb{E}_X[A_{m, k}A_{n, k}] \langle\cos(ms)\cos(ns)\rangle_s \\
    & \qquad - \mathbb{E}_X[A_{m, k}B_{n, k}] \langle\cos(ms)\sin(ns)\rangle_s \\
    & \qquad - \mathbb{E}_X[B_{m, k}A_{n, k}] \langle\sin(ms)\cos(ns)\rangle_s \\
    & \qquad + \mathbb{E}_X[B_{m, k}B_{n, k}] \langle\sin(ms)\sin(ns)\rangle_s\bigg)\\
    & = \alpha_0^2\mathbb{E}_X[A_{0,k}^2] + \sum_{m=1}^M\frac{1}{2}\alpha_m^2\bigg[\mathbb{E}_X[A_{m,k}^2] + \mathbb{E}_X[B_{m,k}^2]\bigg] \\
    & = \mathrm{var}_X(A_{:,k})\sum_{m = 0}^M\alpha_m^2 \\
    &\implies \mathrm{var}_X(A_{:,k}) = \lambda_k \bigg( \sum_{m = 0}^M \alpha_{m}^2\bigg)^{-1}
\end{align*}

Thus, letting $A_{:,k}, B_{:,k} \sim \mathcal{N}\Bigl(0, \lambda_k \Bigl( \sum_{m = 0}^M \alpha_{m}^2\Bigl)^{-1}\Bigr)$, the loops are random, periodic and on average match the first and second moments of the flow.

\section{Guesses based on the physical POD modes}\label{sec:pod_guesses}
The derivation of the loop guesses holds for both the physical space and the latent space. Therefore, it is reasonable to ask whether the autoencoder (AE) step used in this paper is even necessary, and if loop guesses purely based on the physical POD modes are just as good. To compare the two approaches we generate guesses based on the first 3 physical POD modes for the $L=39$ system with $M = 1,2,3,4$ (250 each) with $T_{guess} = 25M$. The table below summarises the success rates for both methods.
\newline

\begin{center}
    \begin{tabular}{|c|c|c|}
        \hline
        $(M,T_{guess})$ & \textbf{Phys. POD} & \textbf{AE + latent POD} \\ \hline
        (1, 25) & 60.0\% & 69.5\% \\ \hline
        (2, 50) & 66.4\% & 76.4\% \\ \hline
        (3, 75) & 22.4\% & 40.6\% \\ \hline
        (4, 100) & 20.8\% & 27.5\% \\ \hline
    \end{tabular}
\end{center}

These results show that the success rates are consistently higher when including the autoencoder. Note also that while the success rates seem to get closer to each other for $(M,T_{guess}) = (4, 100)$, the POD-only method yielded only 12 distinct UPOs with $T>100$ among 250 guesses, while the autoencoder approach yielded 26 unique UPOs with $T>100$ among only 200 guesses. This indicates that not only do the autoencoder guesses converge more often, the UPOs they converge to are also more varied. For $L=100$, we ran a similar test with the first 9 physical POD modes. We generated 250 guesses for each $M=1$ and $M=2$ (the guessed periods correspond approximately to $M=1$ in the autoencoder approach), and only converged to UPOs in 4.2\% of cases, compared to 15.3\% in the autoencoder approach. A significant increase when including the autoencoder.

\section{Supplementary tables}\label{sec:app_tables}

\begin{table*}[htp]
  \centering
  \label{tab:searches_summary}
  \begin{tabular}{|c|cccc|c|c|c|c|c|}
    \hline
        Run & 1 & 2 & 3 & 4 & Fixed points & No convergence & Guesses & Periodic orbits & PO Percentage \\
        \hline
        $T_{guess} = 25, M = 1$ & 139 & 0 & 0 & 0 & 55 & 6 & 200 & 139 & 69.5 \\
        $T_{guess} = 50, M = 2$ & 55 & 327 & 0 & 0 & 70 & 48 & 500 & 382 & 76.4 \\
        $T_{guess} = 75, M = 3$ & 12 & 27 & 103 & 0 & 28 & 180 & 350 & 142 & 40.57 \\
        $T_{guess} = 80, M = 3$ & 7 & 24 & 91 & 0 & 23 & 205 & 350 & 122 & 34.86 \\
        $T_{guess} = 100, M = 4$ & 1 & 8 & 13 & 33 & 10 & 135 & 200 & 55 & 27.5 \\
        $T_{guess} = 105, M = 4$ & 1 & 6 & 12 & 35 & 6 & 140 & 200 & 54 & 27.0 \\
        $T_{guess} = 110, M = 4$ & 2 & 10 & 3 & 39 & 6 & 140 & 200 & 54 & 27.0 \\
        $T_{guess} = 115, M = 4$ & 2 & 5 & 3 & 36 & 11 & 143 & 200 & 46 & 23.0 \\
        $T_{guess} = 120, M = 4$ & 0 & 6 & 1 & 38 & 7 & 148 & 200 & 45 & 22.5 \\
    \hline
    \end{tabular}
  \caption{Summary of periodic orbit searches for $L = 39$. The second column indicates the number of orbits found with Poincaré intersections $p = 1,2,3 $  and 4. One can nicely see that $p$ seems to scale with $M$. Note that the number of periodic orbits found per run is not equal to the number of distinct UPOs found.}
\end{table*}

\renewcommand{\arraystretch}{1.5}
\begin{table*}[htp]
\large
  \centering
  
  \label{tab:gluing_summary}
  \resizebox{\textwidth}{!}{%
  \begin{tabular}{|c|cccccccccccccccccc|}
    \hline
    &24.908&25.371&50.368&52.043&53.135&57.227&57.627&75.281&75.719&75.943&76.621&76.847&76.947&77.37&83.438&83.804&85.536&85.559 \\
    \hline
    24.908&S&50.368&75.281&76.621&N.C.&83.438&83.438&100.189&100.629&100.794&101.506&101.708&101.777&102.68&108.45&108.853&110.385&110.468\\
    25.371&50.368&S&75.719&76.847&N.C.&N.C.&83.804&100.794&101.093&101.253&101.708&102.298&102.355&102.859&109.581&109.969&110.688&N.C.\\
    50.368&75.281&75.943&S&102.126&N.C.&N.C.&108.783&125.656&126.088&126.3&126.933&127.276&127.356&127.913&134.025&134.462&135.791&N.C.\\
    52.043&76.621&76.947&102.187&S&N.C.&N.C.&N.C.&127.123&127.463&127.882&128.633&128.833&N.C.&129.178&134.709&135.244&137.429&136.794\\
    53.135&N.C.&76.847&N.C.&105.787&S&109.637&109.637&N.C.&129.936&N.C.&N.C.&130.734&130.253&130.855&N.C.&N.C.&135.952&N.C.\\
    57.227&83.438&N.C.&N.C.&111.053&N.C.&S&114.867&N.C.&N.C.&N.C.&N.C.&132.15&135.803&133.448&N.C.&N.C.&143.708&N.C.\\
    57.627&83.438&N.C.&N.C.&110.301&N.C.&N.C.&S&N.C.&134.175&N.C.&N.C.&132.15&135.803&N.C.&N.C.&N.C.&143.708&N.C.\\
    75.281&100.189&100.629&125.649&127.077&N.C.&N.C.&133.618&S&151.001&151.217&151.86&152.097&152.168&152.822&158.804&159.207&160.781&160.818\\
    75.719&100.629&101.253&126.126&127.518&N.C.&N.C.&N.C.&151.0&S&151.634&152.286&152.599&152.682&153.16&159.347&159.78&161.291&N.C.\\
    75.943&100.794&101.253&126.3&127.882&N.C.&N.C.&N.C.&151.217&151.634&S&152.62&152.809&152.921&153.492&159.082&159.564&161.549&161.549\\
    76.621&101.506&102.187&126.963&N.C.&130.424&135.819&135.819&151.837&152.315&152.62&S&153.479&153.368&153.89&160.108&160.592&162.239&162.087\\
    76.847&101.708&102.036&127.175&128.414&131.098&N.C.&N.C.&152.181&152.524&152.809&153.194&S&153.691&154.195&161.459&161.865&162.47&N.C.\\
    76.947&101.777&102.036&127.23&128.93&N.C.&N.C.&N.C.&152.149&152.575&152.921&153.583&153.9&S&154.294&N.C.&N.C.&162.169&162.544\\
    77.37&102.68&102.859&127.913&129.178&130.855&133.448&N.C.&152.822&153.16&153.492&153.89&154.195&154.294&S&162.073&N.C.&N.C.&N.C.\\
    83.438&108.45&108.783&133.95&N.C.&132.605&140.689&140.995&158.94&159.313&159.082&160.306&N.C.&160.203&162.073&S&167.234&N.C.&169.078\\
    83.804&108.853&109.085&134.051&N.C.&136.337&141.217&141.465&159.285&159.449&159.564&160.68&N.C.&160.697&N.C.&N.C.&S&169.442&169.467\\
    85.536&110.468&N.C.&N.C.&136.794&N.C.&N.C.&N.C.&160.718&N.C.&161.549&162.087&N.C.&162.544&N.C.&169.078&169.361&S&170.924\\
    85.559&110.468&110.879&135.945&137.429&135.952&143.708&143.708&160.828&161.291&161.549&162.239&162.47&162.274&N.C.&N.C.&169.442&171.096&S\\
    \hline
  \end{tabular}}
  \centering
  \caption{Results of gluing runs. Upper triangular entries: $(i,j)$ corresponds to $\mathcal{P}_i$ glued to $\mathcal{P}_j$. Lower triangular entries: $(i,j)$ corresponds to $\mathcal{P}_i$ glued to $\mathcal{P}_j^s$. N.C. stands for "no convergence", while "S" marks the symmetry line of the matrix.}
\end{table*}

\bibliography{refs}

\end{document}